\documentclass[aps,twocolumn,showpacs,amsmath,amssymb,prc]{revtex4-2}
\usepackage{graphicx,here}
\usepackage{multirow}
\usepackage{tikz}
\usepackage{epstopdf}
\usepackage[percent]{overpic}
\usepackage{scrextend}
\usepackage{xcolor,xspace}
\usepackage[ulem=normalem]{changes}
\usepackage{hyperref}
\hypersetup{colorlinks=True,urlcolor=blue,linkcolor=blue,citecolor=blue,filecolor=black}

\colorlet{Changes@Color}{red}
\usepackage{dcolumn,color,footnote,bm,braket}
\usepackage{url,longtable,tabularx}
\usepackage{threeparttable}

\usepackage{fancybox}
\usetikzlibrary{matrix}

\newcommand\+{\dagger}

\begin{document}

\title{Effect of configuration mixing on quadrupole and octupole collective states of transitional nuclei}

\author{Kosuke Nomura}
\email{knomura@phy.hr}
\affiliation{Department of Physics, Faculty of Science, 
University of Zagreb, HR-10000 Zagreb, Croatia}

\date{\today}

\begin{abstract}
A model is presented that simultaneously describes   
shape coexistence and quadrupole and octupole 
collective excitations within a theoretical framework 
based on the nuclear density functional theory 
and the interacting boson model. 
An optimal interacting-boson Hamiltonian that 
incorporates the configuration mixing between normal and 
intruder states, as well as the octupole degrees of freedom, 
is identified by means of 
self-consistent mean-field calculations using a universal 
energy density functional and a pairing interaction, 
with constraints on the triaxial quadrupole and 
the axially-symmetric quadrupole and octupole 
shape degrees of freedom. 
An illustrative application to the transitional 
nuclei $^{72}$Ge, $^{74}$Se, $^{74}$Kr, and $^{76}$Kr 
shows that the inclusion of the intruder states and 
the configuration mixing 
significantly lower the energy levels of the excited 
$0^+$ states, and that the predicted low-lying positive-parity 
states are characterized by the strong admixture of 
nearly spherical, weakly deformed oblate, and strongly 
deformed prolate shapes. The low-lying 
negative-parity states are shown to be dominated 
by the deformed intruder configurations. 
\end{abstract}

\maketitle

\section{Introduction}

The phenomenon of shape coexistence in  
atomic nuclei has attracted considerable attention 
for many decades \cite{heyde1983,wood1992,andre00,heyde2011}. 
It is observed 
in a wide mass range in the chart of nuclides, 
and is often manifested 
by the appearance of several low-lying excited 
$0^+$ states close in energy to the $0^+$ ground state. 
In a spherical shell model, 
the emergence of the low-energy $0^+$ levels is 
attributed to multiparticle-multihole intruder 
excitations across shell gaps. 
Due to the cross-shell excitations and the 
configuration mixing between the normal 
and intruder states, correlations between 
valence neutrons and protons are enhanced  
to such a degree as to sufficiently lower 
the $0^+$ energies 
\cite{federman1977,duppen1984,heyde1985,heyde1988,heyde92,heyde1995}.
The observed excited $0^+$ intruder states 
can be, in a mean-field picture, 
associated with deformations 
of the intrinsic nuclear shapes, 
as indicated by the competing minima that 
appear close in energy to each other 
in the potential energy surfaces defined in terms 
of the relevant shape degrees of freedom 
\cite{bengtsson1987,bengtsson89,Naza93,andre00,Duguet03,Rayner04Pb,li2016}.

The deformation of the ground-state shape in most 
medium-heavy and heavy nuclei is of reflection 
symmetric, quadrupole type. 
Additional degree of freedom 
that is essential to characterize the nuclear shape 
is the reflection asymmetric, octupole, 
deformation. The octupole correlations are enhanced 
at specific proton $Z$ and neutron $N$ numbers 
often referred to as the ``octupole magic numbers,'' 
34, 56, 88, 134, etc., at which 
the coupling occurs between orbitals with 
opposite parities 
that differ by $\Delta\ell=3\hbar$ and $\Delta j=3\hbar$, 
with $\ell$ and $j$ being, respectively, the orbital 
and total angular momenta of a single nucleon 
\cite{butler1996,butler2016}. 
Typical observables for the octupole collectivity 
are the low-lying negative-parity states, forming 
an approximate alternating-parity 
rotational band with the positive-parity yrast states, 
and the strong electric dipole and octupole 
transitions within the band. 
Empirical evidence for the stable octupole shape  
is mostly concentrated on the axially deformed 
actinides \cite{gaffney2013,chishti2020} and lanthanides 
\cite{bucher2016,bucher2017}. Related theoretical investigations 
have been made from various perspectives 
(see recent reviews, e.g., \cite{butler2016,butler2020b} 
and references are therein).

The octupole collectivity is supposed to be 
present in the transitional nuclei 
in the mass $A \approx$ 70 and 90 regions as well, 
which correspond to the neutron and proton 
``octupole magic numbers'' 
$(N,Z)$ $\approx$ $(34,34)$ and $(56,34)$, 
respectively. 
Together with the octupole collectivity, 
the low-energy nuclear structure in these nuclear systems 
is characterized by a spectacular coexistence between 
prolate and oblate shapes, and a rapid structural evolution 
from one nucleus to another along a given isotopic chain 
(see, e.g., 
Refs.~\cite{clement2007,heyde2011,clement2016,kremer2016,gerst2022,se72oct-2022}, for empirical evidence). 
These facts make it especially attractive, and challenging, 
to explore the transitional nuclei in these regions.

In this paper, a model is presented that simultaneously 
treats the shape coexistence and the quadrupole-octupole 
collective excitations within a theoretical framework 
based on the nuclear density functional theory 
and the interacting boson model (IBM).  
Here the parameters of a version of the IBM, 
that is appropriate for computing spectroscopic observables 
characterizing the shape coexistence, as well as the 
octupole collective states, are determined 
by using the results of self-consistent mean-field (SCMF) 
calculations employing a universal energy density functional (EDF) 
\cite{bender2003,niksic2011,robledo2019}. 
A proof of the method 
is presented in an illustrative application 
to the nuclei 
$^{72}$Ge, $^{74}$Se, $^{74}$Kr, and $^{76}$Kr. 
All these are typical transitional nuclei in the neighbourhood 
of the neutron $N=40$ subshell gap and the proton 
``octupole magic number'' $Z=34$, at which both the shape 
coexistence and octupole correlations are expected to emerge. 
Note that IBM calculations including  
both of these effects were carried out, 
but on purely phenomenological grounds, 
to analyze the intruder and quadrupole-octupole 
coupled states in Cd isotopes 
\cite{Garrett1999-112Cd,lehmann1999,cd114-2003}. 

In the following, a formalism of the configuration-mixing 
IBM framework that includes both the quadrupole and octupole 
degrees of freedom is given (Sec.~\ref{sec:ibm}). 
It is then demonstrated that the parameters of the 
proposed boson model Hamiltonian are obtained 
from the SCMF calculations (Sec.~\ref{sec:map}). 
Spectroscopic observables relevant to the 
shape coexistence and octupole collectivity, including 
the energy levels of the excited $0^+$ states, 
and negative-parity states, and their transition 
properties, are shown in comparison with the experimental data 
(Sec.~\ref{sec:results}). 
Finally, a summary of the main results is given (Sec.~\ref{sec:summary}).

\section{Configuration-mixing $sdf$ IBM\label{sec:ibm}}

The building blocks of the IBM are the monopole $s$, 
quadrupole $d$, and octupole $f$ bosons, 
which represent, from a microscopic 
point of view \cite{OAIT,OAI}, spin and parity $0^+$, 
$2^+$, and $3^-$ pairs of valence nucleons, respectively. 
The number of bosons, $n$, is equal to that of the nucleon 
pairs, and is conserved for each nucleus. 

To incorporate in the IBM system intruder states 
that are associated with 
the shell-model-like 2p-2h, 4p-4h, etc. excitations, 
the boson Hilbert space can be defined as the direct 
sum \cite{duval1981,duval1982}
\begin{align}
\label{eq:space}
 [(sdf)^n]\oplus[(sdf)^{n+2}]\oplus[(sdf)^{n+4}]\oplus\cdots \; ,
\end{align}
where $[(sdf)^{n+2k}]$ ($k=0,1,2,\ldots$) denotes the subspace 
that represents the configuration of $2k$-particle-$2k$-hole 
($2k$p-$2k$h) excitations, comprising 
the $n+2k$ $s$, $d$, and $f$ bosons. 
The basic assumption is that, like in the conventional 
IBM, particlelike and holelike bosons are not distinguished, 
hence the neighboring unperturbed subspaces 
$[(sdf)^{n+2k}]$ differ in boson number by two. 
To avoid complication, no distinction is here made between 
proton and neutron bosons. 
In what follows, short-hand notations 
$n_k \equiv n+2k$, and $[n_k] \equiv [(sdf)^{n+2k}]$ 
are used.

The IBM Hamiltonian that is to carry out the configuration 
mixing of the normal 0p-0h and intruder states is given as
\begin{align}
\label{eq:ham0}
 \hat H 
= &
\,{\hat{P}}_{0} \hat H_{k} {\hat{P}}_{0}
+ \sum_{k=1} 
{\hat{P}}_{k} (\hat H_{k}+\Delta_{k}) {\hat{P}}_{k}
\nonumber\\
 &
\quad\quad
 + \sum_{k=0} {\hat{P}}_{k+1} \hat V_{k,k+1} {\hat{P}}_{k}
   + \textnormal{(H.c.)}
 \; ,
\end{align}
where ${\hat{P}}_{k}$ represents 
a projection operator onto the $2k$p-$2k$h $[n_k]$ configuration, 
$\hat H_{k}$ the corresponding 
unperturbed Hamiltonian, $\Delta_{k}$ the energy 
needed to promote $k$ bosons across the shell closure, 
and $\hat V_{k,k+1}$ the interaction that admixes the 
$[n_k]$ with $[n_{k+1}]$ configurations. 
Couplings between the configuration spaces that 
differ by more than two bosons are not considered, 
since there is no nucleon-nucleon interaction 
that connects such spaces.

The $sdf$-IBM Hamiltonian for each unperturbed space is 
chosen to be of the form
\begin{align}
\label{eq:ham}
 \hat H_{k} = 
&\,\epsilon_{d,k} \hat n_{d} +\epsilon_{f,k} \hat n_{f} 
  + \kappa_{2,k} \hat Q \cdot \hat Q
  + \kappa_{3,k} \hat O \cdot \hat O 
\nonumber\\
&\quad\quad 
  + \rho_k \hat L \cdot \hat L + \eta_k \hat \Theta \; .
\end{align}
In the first (second) term, $\hat n_d=d^\+\cdot\tilde d$ 
($\hat n_f = f^\+ \cdot \tilde f$), 
with $\epsilon_{d,k}$ ($\epsilon_{f,k}$) representing 
the single $d$ ($f$) boson energy relative to the 
$s$-boson one. 
Note $\tilde d_{\mu} = (-1)^{\mu} d_{-\mu}$ 
and $\tilde f_{\mu} = (-1)^{3+\mu} f_{-\mu}$. 
The third and fourth terms in (\ref{eq:ham}) stand for 
quadrupole-quadrupole and octupole-octupole 
interactions, respectively. 
The quadrupole $\hat Q$ and octupole $\hat O$ 
operators read
\begin{align}
& \hat Q = s^\+ \tilde d + d^\+ s 
+ \chi_k (d^\+ \times \tilde d)^{(2)}
+ \chi_k' (f^\+ \times \tilde f)^{(2)} \; , 
\\
& \hat O = s^\+\tilde f + f^\+ s
+ \chi_k''(d^\+\times\tilde f +f^\+\times\tilde d)^{(3)} \; ,
\end{align}
with $\chi_k$, $\chi'_k$, and $\chi''_k$ 
dimensionless parameters. 
The last two terms in (\ref{eq:ham}), 
$\hat L\cdot \hat L$ and $\hat\Theta$, are here 
introduced 
to describe quantitative details of energy spectra. 
The term $\hat L\cdot \hat L$, with $\hat L$ the boson 
angular momentum operator 
$\hat L=\sqrt{10}(d^\+\times\tilde d)^{(1)}$, 
plays a role to describe deformed rotors, 
and is specifically considered for the well deformed 
configurations. 
The term $\hat \Theta$ denotes 
a three-body boson term of the form
\begin{align}
 \hat \Theta = \sum_\lambda 
((d^\+\times d^\+)^{(\lambda)}\times d^\+)^{(3)} \cdot
((\tilde d \times \tilde d)^{(\lambda)}\times \tilde d)^{(3)} \; ,
\end{align}
and is particularly needed to describe the 
quasi-$\gamma$ band of $\gamma$ soft nuclei  
\cite{vanisacker1981,heyde1984,casten1985,nomura2012tri}. 
Note also that the Hamiltonian $\hat H_k$ contains 
two-body $f$-boson interactions in the 
$\hat Q\cdot Q$ and $\hat O\cdot \hat O$ terms. 
This is to account for some observed nonyrast states 
that require more than one $f$ boson in the model space, 
e.g., those of double-octupole phonon nature. 
$\kappa_{2,k}$, $\kappa_{3,k}$, $\rho_k$, and $\eta_k$ 
are interaction strengths. 
The mixing interaction $\hat V_{k,k+1}$ in (\ref{eq:ham0}) 
reads
\begin{align}
\label{eq:mix}
 \hat V_{k,k+1}
&= \omega_{s,k} (s^\+ \cdot s^\+)
 + \omega_{d,k} (d^\+ \cdot d^\+)
\nonumber\\
& \quad\quad 
 + \omega_{f,k} (f^\+ \cdot f^\+)
 + \textnormal{(H.c.)} \; ,
\end{align}
with $\omega_{s,k}$, $\omega_{d,k}$, and $\omega_{f,k}$ 
the strength parameters.

The geometric structure of the IBM is analyzed by 
means of the coherent state introduced 
in Refs.~\cite{dieperink1980,ginocchio1980,bohr1980}. 
In a similar fashion to Eq.~(\ref{eq:space}), 
the coherent state for the configuration mixing IBM, 
$\ket{\Psi}$, is given as a direct sum \cite{frank2004}
\begin{align}
\label{eq:coherent}
 \ket{\Psi} =
 \ket{\Psi_{0}(\vec\alpha_0)} \oplus 
 \ket{\Psi_{1}(\vec\alpha_1)} \oplus
 \ket{\Psi_{2}(\vec\alpha_2)} \oplus \cdots \; .
\end{align}
$\ket{\Psi_k(\vec\alpha_k)}$ 
stands for the coherent state for each unperturbed $[n_k]$ 
boson subspace, and is defined, up to the normalization factor, 
as
\begin{align}
 \ket{\Psi_{k}(\vec\alpha_{k})}
\propto
\left(
s^\+ 
+ \sum_\mu \alpha_{2\mu}^{(k)} d^\+_{\mu}
+ \sum_\mu \alpha_{3\mu}^{(k)} f^\+_{\mu}
\right)^{n_k}
\ket{0} \; .
\end{align}
The ket $\ket{0}$ denotes the boson 
vacuum, i.e., the inert core. 
Real coefficients 
$\vec\alpha_{k}\equiv(\alpha_{2\mu}^{(k)},\alpha_{3\mu}^{(k)})$ 
represent amplitudes analogous to the 
collective variables in the geometrical 
model \cite{BM}. Specifically, 
\begin{align}
 \alpha^{(k)}_{20}=\beta_{2,k}\cos\gamma_k\;,\quad
 \alpha^{(k)}_{2\pm1}=0\;,\quad
 \alpha^{(k)}_{2\pm2}=\frac{1}{\sqrt{2}}\beta_{2,k}\sin\gamma_k \; ,
\end{align}
while, for the octupole modes, only the axially symmetric 
deformation is assumed to be relevant, i.e., 
\begin{align}
\alpha^{(k)}_{3\mu}=\beta_{3,k}\delta_{\mu,0} \; . 
\end{align}
It is further assumed that the amplitudes $\beta_{2,k}$ 
and $\beta_{3,k}$ are proportional to the axial quadrupole 
and octupole deformations in the geometrical model, i.e., 
\begin{align}
\label{eq:c23}
\beta_{2,k}\equiv C_{2,k} \beta_2 \; , 
\quad
\beta_{3,k}\equiv C_{3,k} \beta_3 \; ,
\end{align}
with constants of 
proportionality $C_{2,k}$ and $C_{3,k}$ defined 
separately for the unperturbed spaces, and that the 
triaxiality $\gamma_k$ in the boson system is identical to the 
fermionic counterpart, $\gamma$, and is common for all the 
unperturbed spaces, i.e., $\gamma_k = \gamma$.

The potential energy surface for the boson system is 
defined in terms of the 
three deformation variables $\beta_2$, $\gamma$, and $\beta_3$, 
and is calculated as the expectation value of the Hamiltonian 
(\ref{eq:ham0}) in the coherent state of (\ref{eq:coherent}). 
The energy surface is actually obtained in the form of 
a $(k+1)\times (k+1)$ matrix ${\bf E}(\beta_2,\gamma,\beta_3)$. 
The diagonal element of the matrix is calculated as
\begin{align}
\label{eq:pes-diag}
 E_{k,k}(\beta_2,\gamma,\beta_3)
&=\braket{\Psi_k(\vec\alpha_k)|\hat H_k|\Psi_k(\vec\alpha_k)}
+\Delta_k \; ,
\end{align}
where $\Delta_k$ enters only 
for $k\geqslant1$, while the nondiagonal part 
\begin{align}
\label{eq:pes-nondiag}
  E_{k,k+1}(\beta_2,\gamma,\beta_3)
=&\,E_{k+1,k}(\beta_2,\gamma,\beta_3)
\nonumber\\
=&\braket{\Psi_{k+1}(\vec\alpha_{k+1})|\hat V_{k,k+1}|\Psi_k(\vec\alpha_k)}
\; .
\end{align}
Explicit analytical forms of the quantities 
(\ref{eq:pes-diag}) and (\ref{eq:pes-nondiag}) are
given in the Appendix~\ref{sec:pes}. 
As in the literature \cite{frank2004,frank2006}, 
the lowest-energy eigenvalue of the coherent-state 
matrix ${\bf E}(\beta_2,\gamma,\beta_3)$ 
at each set of the coordinates $(\beta_2,\gamma,\beta_3)$ 
is here taken as the energy surface for the boson system.

The bosonic energy surface depends on the 
parameters of the Hamiltonian, and on the 
constants of proportionality $C_{2,k}$ and 
$C_{3,k}$ (\ref{eq:c23}). 
For each $[n_k]$ configuration, there are 
nine parameters 
in the unperturbed Hamiltonian $\hat H_k$, 
the proportionality parameters 
$C_{2,k}$ and $C_{3,k}$ (\ref{eq:c23}), 
and the energy offset $\Delta_k$ ($k\geqslant1$). 
In addition, the mixing interaction $\hat V_{k,k+1}$, 
connecting the subspaces 
$[n_k]$ and $[n_{k+1}]$, has three parameters 
for each $k$ [see Eq.~(\ref{eq:mix})].  
These parameters are determined by the 
procedure described below.

\section{Procedure to build the IBM Hamiltonian\label{sec:map}}

\begin{figure}
\begin{center}
\includegraphics[width=\linewidth]{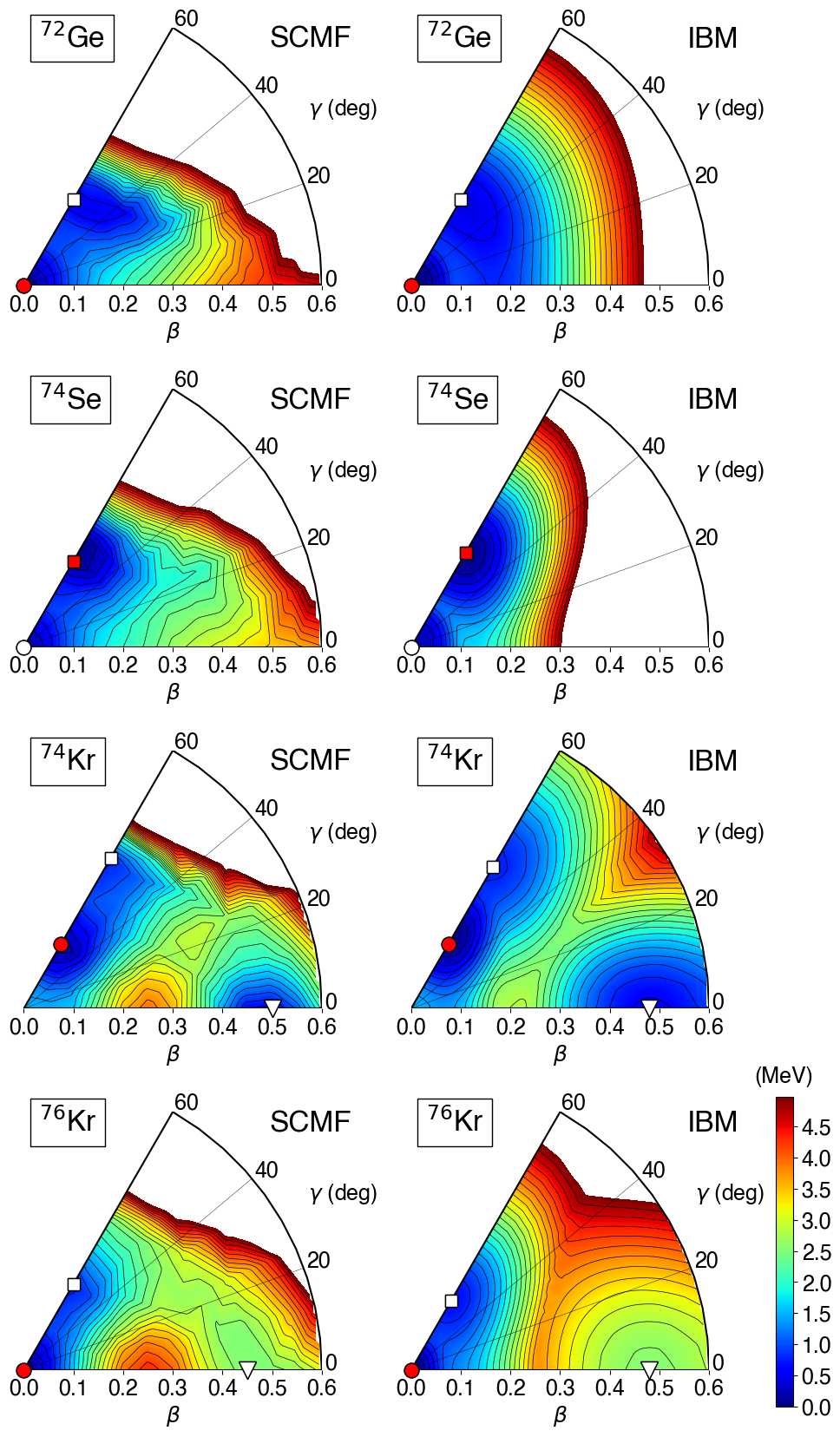}
\caption{Left column: Triaxial quadrupole SCMF 
potential energy surfaces for the $^{72}$Ge, $^{74}$Se, $^{74}$Kr, 
and $^{76}$Kr isotopes 
as functions of the $\beta_2$ and $\gamma$ deformations, 
computed by the constrained relativistic Hartree-Bogoliubov 
method with the density-dependent point-coupling 
interaction and the separable pairing force of finite range. 
Right column: Corresponding energy surfaces for the boson system. 
The energy difference between neighboring contours is 0.2 MeV. 
The minima associated with the 0p-0h, 2p-2h, and 4p-4h 
unperturbed configurations are identified by the circle, 
square, and triangle, respectively, and 
the solid symbol with color red denotes the global minimum. 
}
\label{fig:pes-bg}
\end{center}
\end{figure}

\begin{figure*}
\begin{center}
\includegraphics[width=\linewidth]{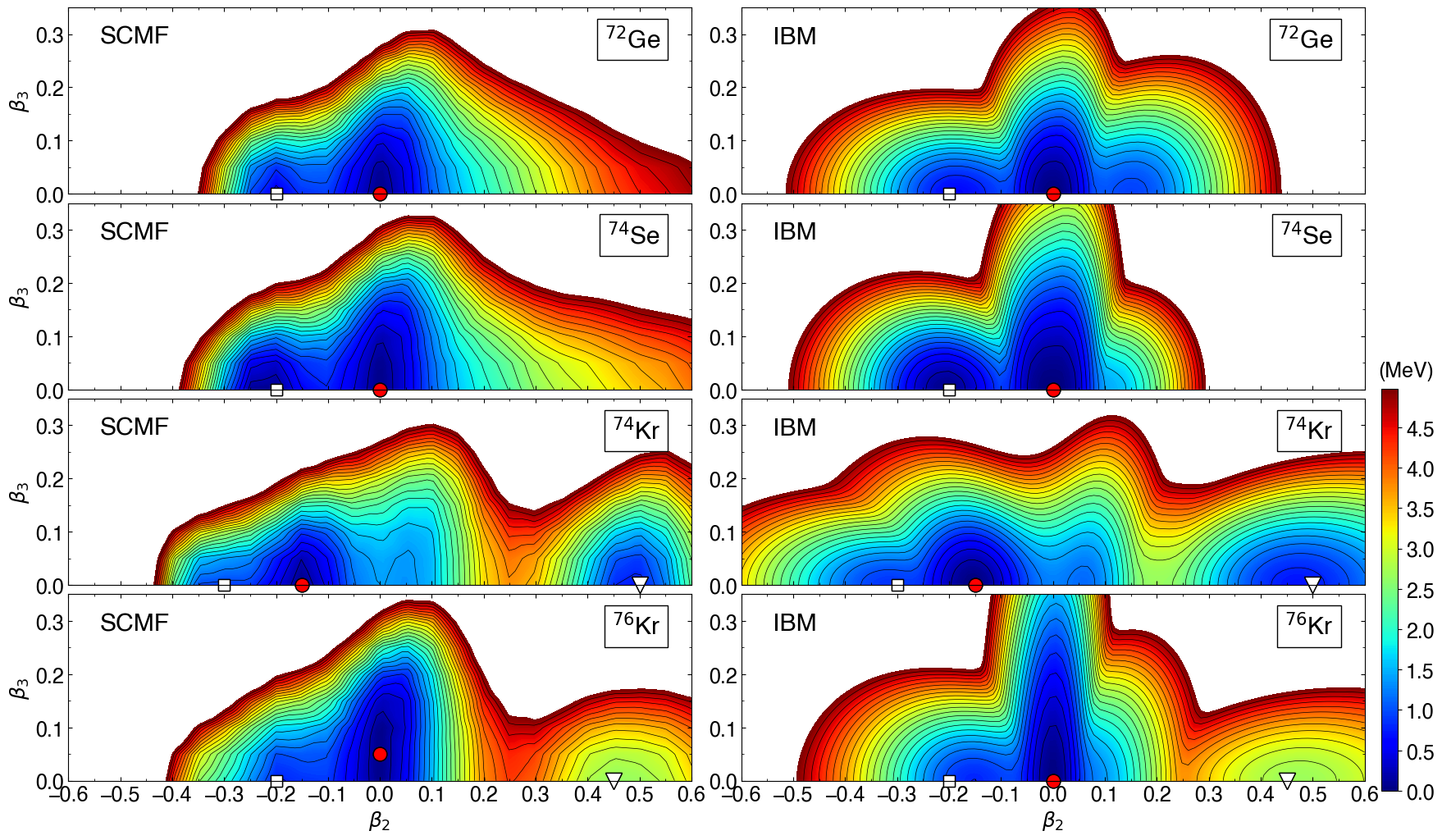}
\caption{Same as Fig.~\ref{fig:pes-bg}, but for the 
axially symmetric quadrupole and octupole SCMF and IBM
potential energy surfaces 
as functions of the $\beta_2$ and $\beta_3$ deformations. 
}
\label{fig:pes-qo}
\end{center}
\end{figure*}

In the first step, for each nucleus, the 
constrained SCMF calculations \cite{RS} are 
carried out within the framework of the relativistic 
Hartree-Bogoliubov method \cite{vretenar2005,niksic2011,DIRHB} 
using the density-dependent point-coupling  
interaction \cite{DDPC1} and the separable 
pairing force of finite range \cite{tian2009}. 
Two sets of the SCMF calculations are performed, 
with constraints on 
the (i) triaxial quadrupole $Q_{20}$ and $Q_{22}$ 
and on the (ii) axial quadrupole $Q_{20}$ and octupole $Q_{30}$ 
moments. 

The SCMF calculations provide potential energy surfaces 
as functions of the triaxial quadrupole 
($\beta_2$-$\gamma$) and axially symmetric 
quadrupole-octupole ($\beta_2$-$\beta_3$) 
deformations, which are shown in the first 
columns of Figs.~\ref{fig:pes-bg} 
and \ref{fig:pes-qo}, respectively. 
Here, the $\beta_2$, $\gamma$, 
and $\beta_3$ deformations can be obtained through the 
relations
\begin{align}
 &\beta_2=\frac{\sqrt{5\pi}}{3 A R_0^2}
\sqrt{{\braket{\hat Q_{20}}}^2 + 2{\braket{\hat Q_{22}}}^2 } \; , \\
 &\gamma = \arctan\sqrt{2}
\frac{\braket{\hat Q_{22}}}{\braket{\hat Q_{20}}} \; ,\\
 &\beta_3=\frac{\sqrt{7\pi}}{3 A R_0^3}\braket{\hat Q_{30}} \; ,
\end{align}
with $R_0=1.2A^{1/3}$ fm.

One finds, in the SCMF $\beta_2$-$\gamma$ energy surfaces 
for $^{72}$Ge and $^{74}$Se (Fig.~\ref{fig:pes-bg}), 
two minima that are close in energy to each other, 
one spherical and the other oblate. 
For $^{74}$Kr ($^{76}$Kr), 
in addition to two oblate (a spherical and an oblate) minima, 
a third, strongly prolate deformed minimum appears 
at the deformation $\beta_2=0.5$ (0.45). 
The prolate minimum in $^{74}$Kr is deeper 
in energy than the one for $^{76}$Kr, and is even close 
to the oblate global minimum.

The $\beta_2$-$\beta_3$ SCMF energy surfaces, 
shown in Fig.~\ref{fig:pes-qo}, exhibit 
$\beta_3$ softness in $^{74}$Se and $^{76}$Kr. 
The former nucleus is predicted to be soft 
in the $\beta_2$ deformation as well. 
The softness implies sizable correlations 
arising from the shape mixing, which play an 
important role in the spectroscopic properties. 
For $^{74}$Se, the global minimum 
in the $\beta_2$-$\beta_3$ deformation 
plane occurs at the spherical 
configuration, while in the $\beta_2$-$\gamma$ plane 
the oblate global minimum is obtained. 
The discrepancy is due to the fact that the constrained 
SCMF calculations are here performed separately within 
the $\beta_2$-$\gamma$ and $\beta_2$-$\beta_3$ deformation 
spaces. However, the energy difference between the two 
minima is negligibly small on both the 
$\beta_2$-$\gamma$ and $\beta_2$-$\beta_3$ surfaces, 
i.e., the spherical local (global) minimum 
in the $\beta_2$-$\gamma$ ($\beta_2$-$\beta_3$) energy surface 
of $^{74}$Se is only 72 (29) keV above (below) the 
oblate global (local) minimum. Such a difference 
would not affect much the resulting spectroscopic 
properties.

The SCMF results are subsequently 
used as microscopic inputs to build the 
configuration-mixing $sdf$-IBM Hamiltonian (\ref{eq:ham0}). 
This procedure consists of a fermion-to-boson mapping 
developed in Refs.~\cite{nomura2008,nomura2010}:
The potential energy surface, 
computed by the constrained SCMF method for 
each nucleus, is mapped 
onto the expectation value of the IBM 
Hamiltonian in the boson coherent state, 
and this mapping procedure specifies the 
strength parameters of the IBM Hamiltonian. 
In other words, the IBM parameters are 
calibrated so that the SCMF and IBM energy surfaces 
are similar in topology to each other. 
The mapping procedure has been extended further 
so as to include in the usual boson space 
the configuration mixing between normal and intruder 
states \cite{nomura2012sc}, and octupole degrees 
of freedom \cite{nomura2014}.

For the considered nuclei, 
the neutron $N=$ 28-50 and proton 
$Z=$ 28-50 major shells are taken as the normal 
configuration space. 
The intruder states are here assumed to be the 
proton excitations across the $Z=28$ major shell. 
The two configurations, $[n_0]$ and $[n_1]$, 
are considered for $^{72}$Ge and $^{74}$Se, since 
the SCMF potential energy surfaces show two minima 
(see Figs.~\ref{fig:pes-bg} and \ref{fig:pes-qo}). 
On the other hand, since there appears an additional 
prolate third minimum for $^{74}$Kr and $^{76}$Kr, 
the three configurations, $[n_0]$, $[n_1]$, 
and $[n_2]$, are included in the boson spaces 
for these nuclei. 
According to the prescription proposed 
in Ref.~\cite{nomura2012sc}, 
the unperturbed Hamiltonians $\hat H_k$ for the 
0p-0h, 2p-2h, and 4p-4h configurations 
are associated with those minima on the SCMF 
$\beta_2$-$\gamma$ energy surface that correspond 
to the smallest, second larger, and third larger 
$\beta_2$ deformations, respectively. 
The correspondence between the $[n_k]$-boson 
configuration and the mean-field minimum is indicated 
in the SCMF energy surfaces in Figs.~\ref{fig:pes-bg} 
and \ref{fig:pes-qo}.

Since the full Hamiltonian (\ref{eq:ham0}) contains 
a large number of parameters, it is plausible to consider 
each sector of the Hamiltonian separately in the following way. 

\begin{enumerate}

\item
First, the $sd$-boson sector of each 
unperturbed Hamiltonian $\hat H_k$ is fixed. 
The relevant parameters, $\epsilon_{d,k}$, 
$\kappa_{2,k}$, $\chi_{k}$, $\eta_k$, and $C_{2,k}$, 
are then chosen so that the topology of the 
$\beta_2$-$\gamma$ SCMF 
energy surface in the vicinity of the 
corresponding minimum is reproduced by the diagonal 
matrix element $E_{k,k}(\beta_2,\gamma,0)$ 
(\ref{eq:pes-diag}), i.e., 
the expectation value of the unperturbed Hamiltonian 
in the $[n_k]$ space. 

\item
The parameter for the $\hat L \cdot \hat L$ term, $\rho_k$, 
is derived separately from those described 
above, in such a way \cite{nomura2011rot} that 
the cranking moment of inertia 
calculated at each mean-field minimum 
in the intrinsic frame of the boson system 
\cite{schaaser1986}, 
is equal to the Inglis-Belyaev \cite{inglis1956,belyaev1961} 
moment of inertia, ${\mathcal{I}}_\text{IB}$, computed 
by the same constrained SCMF method. 
This term is here taken into account in 
those well-deformed configurations for which the 
Inglis-Belyaev values are calculated to be 
${\cal{I}}_\text{IB}>10$ MeV$^{-1}$, specifically, 
the 2p-2h and 4p-4h configurations for $^{74}$Kr, 
and the 4p-4h configuration for $^{76}$Kr. 
It is also noted that the Inglis-Belyaev moments of inertia 
are here increased by 30 \%. 
This is to take into account the fact that the 
Inglis-Belyaev formula gives the moments of inertia that are 
by typically 30-40 \% smaller than the empirical ones. 

\item
The mixing interaction $\hat V_{k,k+1}$ is introduced. 
Here, the strengths $\omega_{s,k}$, 
$\omega_{d,k}$, and $\omega_{f,k}$ are assumed 
to be equal to each other for each $k$, i.e., 
$\omega_{f,k}=\omega_{d,k}=\omega_{f,k}\equiv \omega_k$. 
The energy offset $\Delta_k$, and the 
parameter $\omega_k$ for $\hat V_{k,k+1}$ 
are obtained so that the energy difference between the 
neighboring mean-field minima corresponding to the 
$[n_k]$ and $[n_{k+1}]$ configurations, and the 
topology of the barrier separating the two minima, respectively, 
are reproduced by the lowest-energy eigenvalue 
of the matrix ${\mathbf{E}}(\beta_2,\gamma,0)$. 

\item
Finally, the $f$-boson sector of the Hamiltonian 
(\ref{eq:ham}) is fixed. The relevant parameters, 
$\epsilon_{f,k}$, $\chi_k'$, $\kappa_{3,k}$, $\chi''_k$, and 
$C_{3,k}$, are determined so that the 
diagonal matrix element for the $[n_k]$ subspace, 
$E_{k,k}(\beta_2,0^\circ,\beta_3)$, 
should reproduce the topology of 
the SCMF $\beta_2$-$\beta_3$ energy surface 
in the vicinity of the corresponding minimum. 
In this step, the parameters for the $sd$-boson sector 
of the Hamiltonian, mixing strength $\omega_k$, and 
the offset energy $\Delta_k$, 
obtained in the previous three steps, are kept unchanged. 

\end{enumerate}
Table~\ref{tab:para} lists the adopted parameters 
for the studied nuclei.

\begin{table*}
\caption{
\label{tab:para}
Adopted parameters for the configuration-mixing 
$sdf$-IBM Hamiltonian. See Eqs.~(\ref{eq:ham}) 
and (\ref{eq:mix}), and (\ref{eq:c23}) for definitions. 
}
 \begin{center}
 \begin{ruledtabular}
 \begin{tabular}{cccccccccccccccc}
  & \multirow{2}{*}{Config.} & \multirow{2}{*}{$n_k$} & $\epsilon_{d,k}$ & $\kappa_{2,k}$ & \multirow{2}{*}{$\chi_k$} & \multirow{2}{*}{$\chi_k'$} & $\rho_k$ 
  & $\eta_k$ & $\epsilon_{f,k}$ & $\kappa_{3,k}$ & \multirow{2}{*}{$\chi_k''$} & \multirow{2}{*}{$C_{2,k}$} & \multirow{2}{*}{$C_{3,k}$} & $\omega_k$ & $\Delta_k$ \\
  & & & (MeV) & (keV) & & & (keV) & (keV) & (MeV) & (keV) & & & & (keV) & (MeV) \\
\hline\multirow{2}{*}{$^{72}$Ge} 
  & 0p-0h &  7 & 1.9 & $-85$ & 0.15 & $-0.15$ & 0 
               & 0 & $-2.0$ & 35 & $-0.8$ & 6.5 & 1.9 & & \\
  & 2p-2h &  9 & 1.05 & $-70$ & 0.20 & $-0.50$ & 0 
               & 55 & $-2.2$ & 28 & $-1.6$ & 2.8 & 2.4 & 40 & 1.4\\
[1.0ex]\multirow{2}{*}{$^{74}$Se} 
  & 0p-0h &  8 & 1.9 & $-85$ & 0.15 & $0.15$ & 0 
               & 0 & $-2.0$ & 35 & $-0.8$ & 6.5 & 1.4 & & \\
  & 2p-2h & 10 & 1.2 & $-70$ & 0.67 & $-0.70$ & 0 
               & 50 & $-2.2$ & 28 & $-1.4$ & 3.2 & 2.4 & 35 & 1.9\\
[1.0ex]\multirow{3}{*}{$^{74}$Kr} 
  & 0p-0h &  9 & 1.0 & $-57$ & 0.95 & $0.95$ & 0 
               & 0 & $-2.0$ & 25 & $-1.6$ & 4.5 & 2.0 & & \\
  & 2p-2h & 11 & 0.81 & $-56$ & 0.65 & $0.65$ & 6.4 
               & 0 & $-2.0$ & 24 & $-1.4$ & 2.3 & 2.4 & 40 & 2.3\\
  & 4p-4h & 13 & 0.48 & $-42$ & $-0.90$ & $-0.36$ & 12 
               & 0 & $-2.0$ & 23 & $-1.8$ & 2.0 & 2.6 & 30 & 4.7\\
[1.0ex]\multirow{3}{*}{$^{76}$Kr} 
  & 0p-0h &  9 & 1.9 & $-36$ & 0.67 & $0.67$ & 0 
               & 0 & $-2.0$ & 22 & $-0.8$ & 6.0 & 0.9 & & \\
  & 2p-2h & 11 & 0.85 & $-33$ & 0.75 & $0.75$ & 0 
               & 0 & $-2.0$ & 22 & $-1.8$ & 3.6 & 2.2 & 40 & 1.3\\
  & 4p-4h & 13 & 0.42 & $-35$ & $-0.60$ & $-0.60$ & 22 
               & 0 & $-2.1$ & 20 & $-2.0$ & 1.7 & 2.6 & 30 & 5.6\\
 \end{tabular}
 \end{ruledtabular}
 \end{center}
\end{table*}

The $\beta_2$-$\gamma$ 
and $\beta_2$-$\beta_3$ mapped-IBM 
energy surfaces are shown in the right columns 
of Figs.~\ref{fig:pes-bg} 
and \ref{fig:pes-qo}, respectively. 
They are similar in topology 
to the original SCMF energy surfaces up to a few MeV 
excitation from the global minimum. 
There are, however, some discrepancies as well. 
In particular, the SCMF $\beta_2$-$\beta_3$ energy 
surface for $^{76}$Kr shows the global 
minimum at $\beta_3\approx 0.05 \neq 0$. 
The nonzero $\beta_3$ minimum is not 
reproduced in the bosonic one, which rather 
gives the minimum at $\beta_3=0$. 
Note, however, that 
the SCMF energy surface is also considerably 
soft along the $\beta_3$ direction. Indeed the 
global minimum at $(\beta_2,\beta_3)\approx (0.0, 0.05)$ 
is by only 17 keV lower in energy than that 
configuration on the $\beta_3=0$ axis with the 
same $\beta_2$, i.e., 
$(\beta_2,\beta_3)\approx (0.0, 0.0)$. 
The energy difference is so small that it would not 
have much influence on the final results. 

Another discrepancy is that 
the $\beta_2$-$\gamma$ 
SCMF energy surfaces for $^{72}$Ge and $^{74}$Se are 
flat on the prolate side, extending to the region 
corresponding to large $\beta_2$ deformation, while the 
bosonic surfaces are more rigid in $\gamma$ deformation. 
This is partly because 
the analytical form of the IBM energy surface 
has too limited degrees of freedom to reproduce 
such a topology. However, the 
mean-field configurations most relevant 
to the low-lying collective 
states are those in the vicinity of each minimum, 
hence the mapping is carried out primarily 
to reproduce the topology around the minimum.

\section{Results for the spectroscopic properties\label{sec:results}}

With all the parameters determined by the 
procedure described in the previous section, 
the Hamiltonian (\ref{eq:ham0}) is diagonalized 
within the boson Hilbert space defined in Eq.~(\ref{eq:space}), 
producing spectroscopic observables. 
For the diagonalization, the boson $m$-scheme basis 
is used. 

In order to reduce computational time, 
a truncation is made 
so that the maximum number of $f$ bosons is limited 
as $n_f^\text{max}=3$. 
The truncation makes sense, particularly because the 
majority of the low-energy states with positive and 
negative parity turn out 
to be essentially of zero- and one-$f$-boson characters, 
respectively. 
Sensitivity of the calculated observables 
to this truncation will be discussed in detail 
in Sec.~\ref{sec:fboson}.

\begin{figure*}
\begin{center}
\includegraphics[width=.49\linewidth]{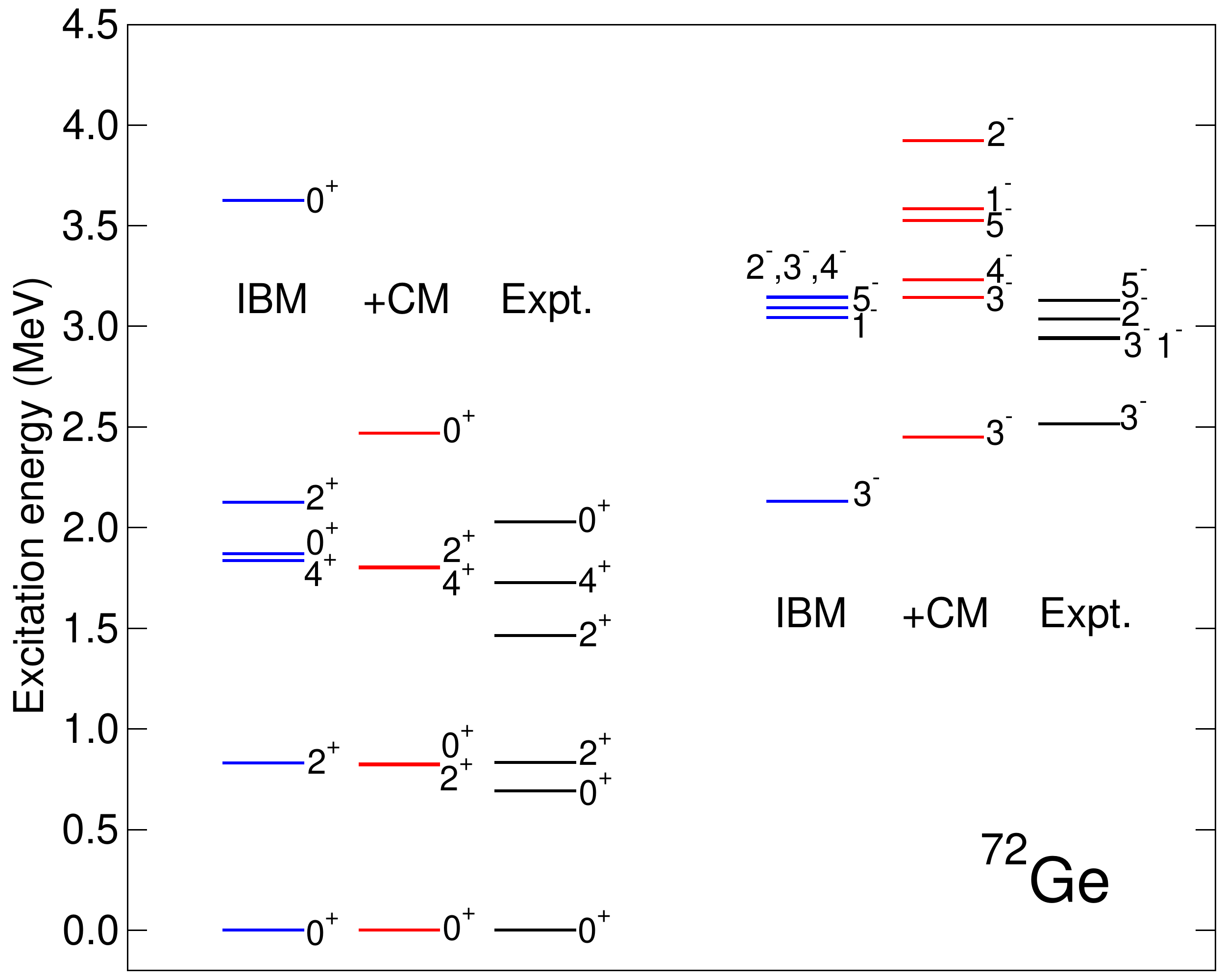}
\includegraphics[width=.49\linewidth]{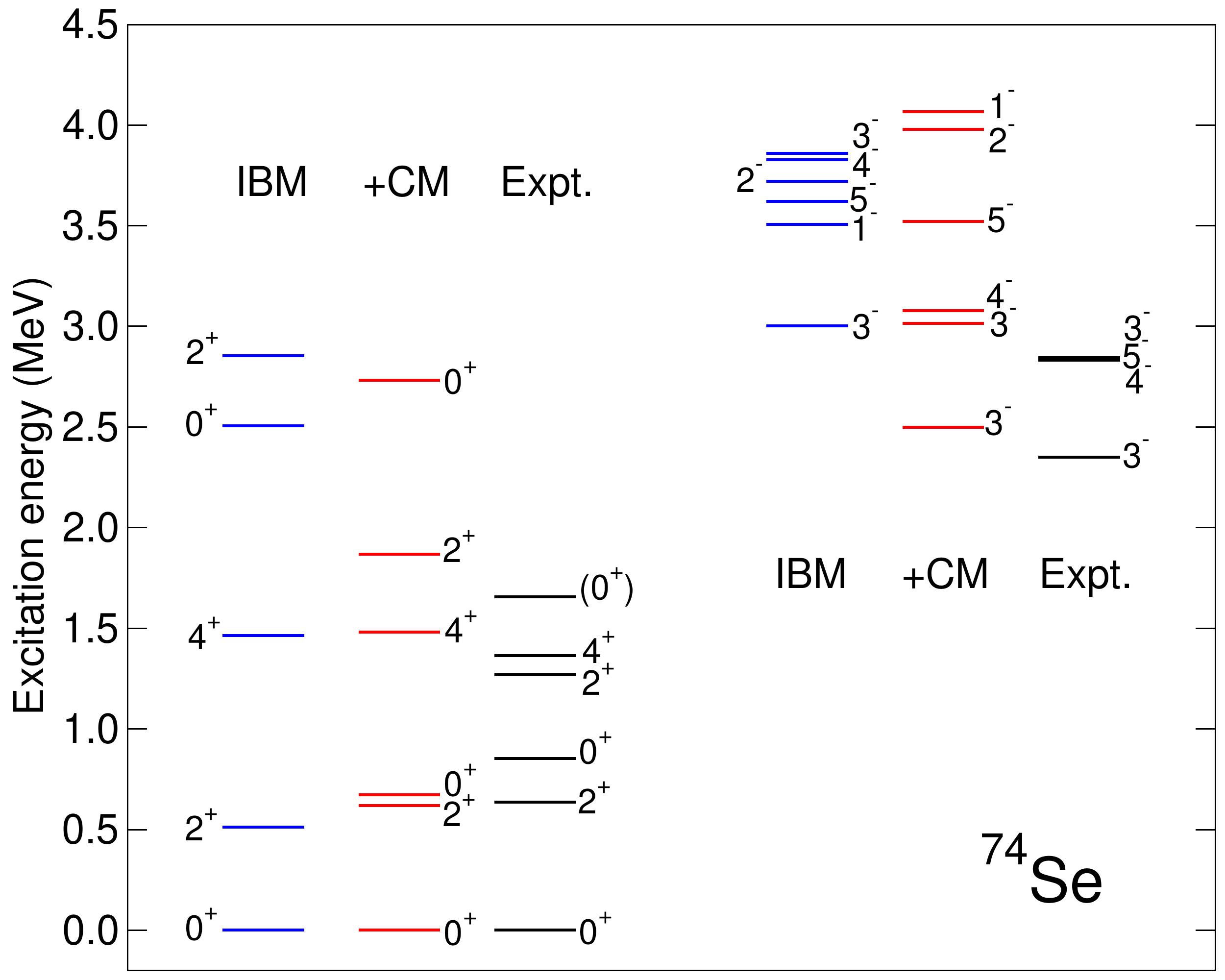}\\
\includegraphics[width=.49\linewidth]{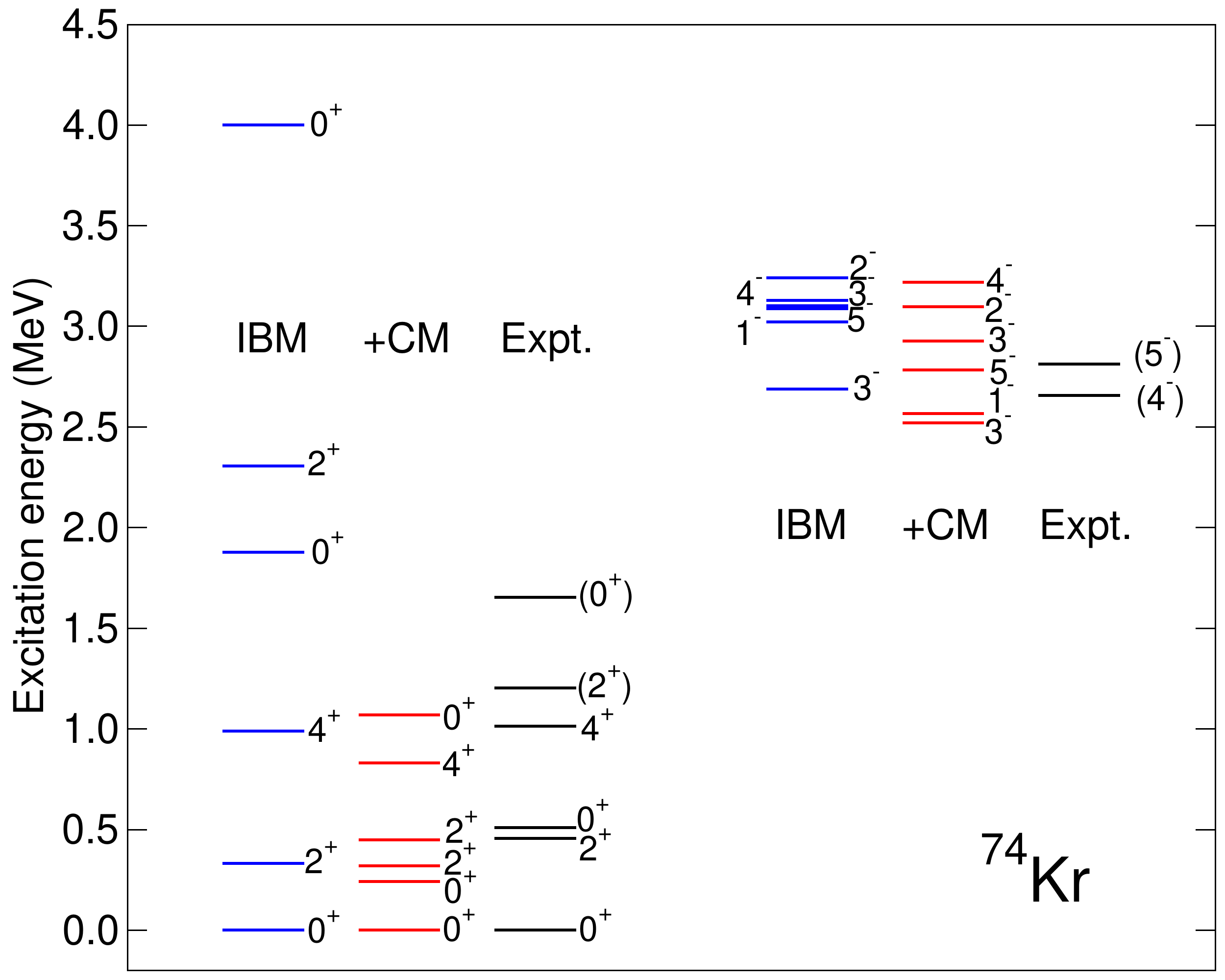}
\includegraphics[width=.49\linewidth]{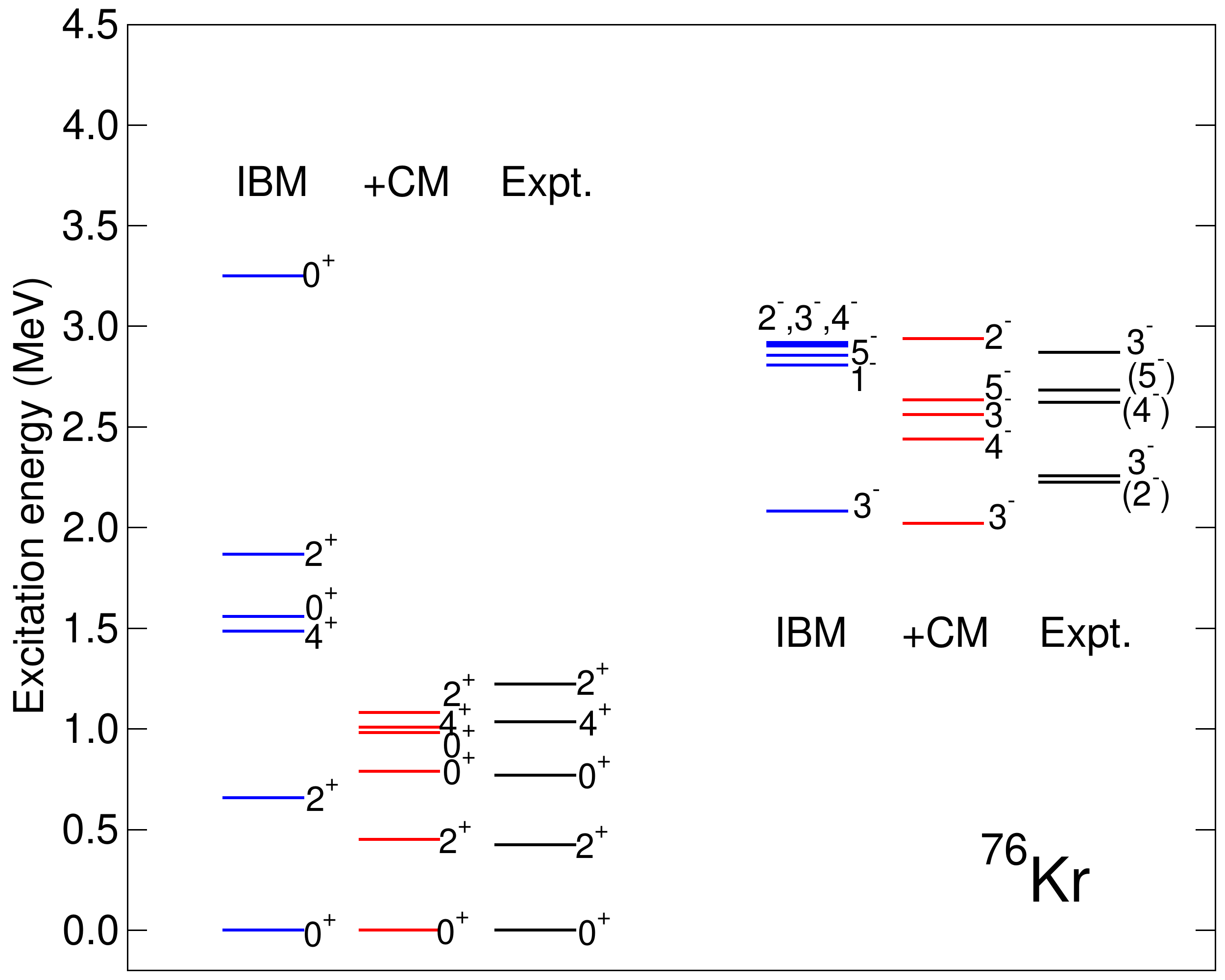}
\caption{Comparison of low-energy excitation 
spectra of positive- and negative-parity states 
calculated with the $sdf$-IBM excluding (``IBM'') 
and including (``+CM'') the configuration mixing. 
Experimental spectra \cite{data} are also shown as 
a reference.}
\label{fig:ibmcm}
\end{center}
\end{figure*}

\subsection{Effect of configuration mixing}

A drastic, as well as favorable, effect of including 
the configuration mixing in the $sdf$-IBM 
is the lowering of the excited $0^+$ states. 
Figure~\ref{fig:ibmcm} 
compares between the low-energy spectra calculated 
within the $sdf$-IBM including and excluding 
the configuration mixing. In the latter calculation, 
only a single configuration corresponding to the $[n_0]$ 
normal space is considered, and the Hamiltonian $\hat H_0$ 
is associated with the global minimum.

By the inclusion of the configuration mixing, 
the $0^+_2$ and $0^+_3$ energy levels are significantly 
lowered in all the four nuclei considered. 
Another non-yrast level, $2^+_2$, is also lowered 
in the mixing calculation. 
The positive-parity yrast states, $2^+_1$ and 
$4^+_1$, remain unchanged for $^{72}$Ge and $^{74}$Se, 
but are lowered for $^{74}$Kr and $^{76}$Kr by 
the mixing. 
The configuration mixing also has impacts on the 
negative-parity levels. 
The $sdf$-IBM calculation with only a single 
configuration generally gives an approximate 
degeneracy of the 
$1^-_1$, $2^-_1$, $3^-_2$, $4^-_1$, and $5^-_1$ levels, 
which is considered a quintet of the 
quadrupole-octupole phonon coupling $2^+ \otimes 3^-$. 
The degeneracy is removed after the mixing, 
as a consequence of the repulsion 
among the low-spin levels.

To interpret the nature of the low-lying states, 
it is useful to analyze fraction of the $2k$p-$2k$h 
components in their wave functions. 
The corresponding results for some low-lying states 
are listed in Table~\ref{tab:frac}. 
Generally, high degree of the mixing appears to be present, 
particularly in the two lowest $0^+$ states. 
The intruder 2p-2h configuration dominates over the normal 
configuration in the yrast states of both parities 
for the $^{72}$Ge, $^{74}$Ge, and $^{76}$Kr nuclei. 
The 4p-4h configuration associated with the strongly 
deformed prolate minimum makes a particularly large contribution 
to the low-lying states of $^{74}$Kr. 

\begin{table}
\caption{
Fractions (in \%) of the $2k$p-$2k$h $[n_k]$ 
configurations in the $sdf$-IBM wave functions 
of low-lying states. 
\label{tab:frac}
}
 \begin{center}
 \begin{ruledtabular}
  \begin{tabular}{llcccc}
 $I^\pi$ & Config. & $^{72}$Ge & $^{74}$Se & $^{74}$Kr & $^{76}$Kr \\
\hline\multirow{3}{*}{$0^+_1$}
& $[n_0]$ & 38 & 37 & 25 & 24 \\
& $[n_1]$ & 62 & 63 & 43 & 71 \\
& $[n_2]$ &    &    & 32 &  5 \\
[1.0ex]\multirow{3}{*}{$0^+_2$}
& $[n_0]$ & 62 & 62 & 19 & 51 \\
& $[n_1]$ & 38 & 38 & 14 & 15 \\
& $[n_2]$ &    &    & 67 & 34 \\
[1.0ex]\multirow{3}{*}{$0^+_3$}
& $[n_0]$ &  0.01 &  3 & 56 & 14 \\
& $[n_1]$ & 99.99 & 97 & 42 & 25 \\
& $[n_2]$ &    &    &  2 & 61 \\
[1.0ex]\multirow{3}{*}{$2^+_1$}
& $[n_0]$ &  8 &  6 & 12 &  4 \\
& $[n_1]$ & 92 & 94 & 24 & 90 \\
& $[n_2]$ &    &    & 63 &  6 \\
[1.0ex]\multirow{3}{*}{$2^+_2$}
& $[n_0]$ &  4 & 84 & 28 &  0 \\
& $[n_1]$ & 96 & 16 & 36 & 18 \\
& $[n_2]$ &    &    & 36 & 82 \\
[1.0ex]\multirow{3}{*}{$4^+_1$}
& $[n_0]$ &  2 &  2 &  1 &  2 \\
& $[n_1]$ & 98 & 98 &  5 & 94 \\
& $[n_2]$ &    &    & 94 &  4 \\
[1.0ex]\multirow{3}{*}{$3^-_1$}
& $[n_0]$ & 28 & 19 &  1.8 & 19 \\
& $[n_1]$ & 72 & 81 &  6.5 & 75 \\
& $[n_2]$ &    &    & 91.7 &  6 \\
[1.0ex]\multirow{3}{*}{$4^-_1$}
& $[n_0]$ &  4 &  3 & 24 &  3 \\
& $[n_1]$ & 96 & 97 & 39 & 92 \\
& $[n_2]$ &    &    & 38 &  5 \\
[1.0ex]\multirow{3}{*}{$5^-_1$}
& $[n_0]$ &  5.8 &  4 & 0.32 &  4 \\
& $[n_1]$ & 94.2 & 96 & 2.44 & 88 \\
& $[n_2]$ &    &    & 97.24 & 8 \\
  \end{tabular}
 \end{ruledtabular}
 \end{center}
\end{table}

\subsection{Detailed energy levels}

\begin{figure}
\begin{center}
\includegraphics[width=\linewidth]{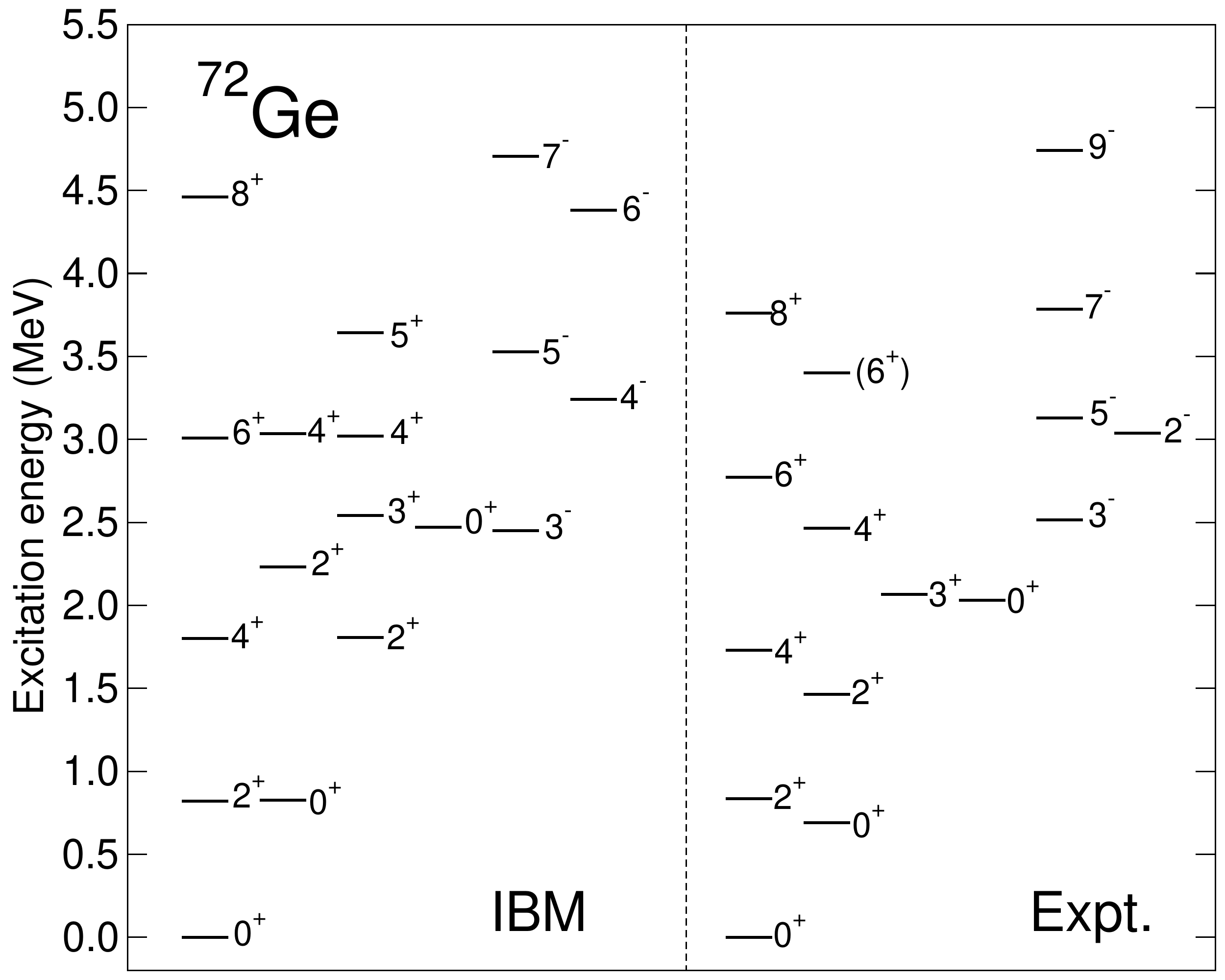}
\caption{Low-energy excitation 
spectra of positive- and negative-parity states of $^{72}$Ge 
calculated with the configuration mixing $sdf$-IBM in 
comparison with experimental data \cite{data}. }
\label{fig:ge72}
\end{center}
\end{figure}

\begin{figure}
\begin{center}
\includegraphics[width=\linewidth]{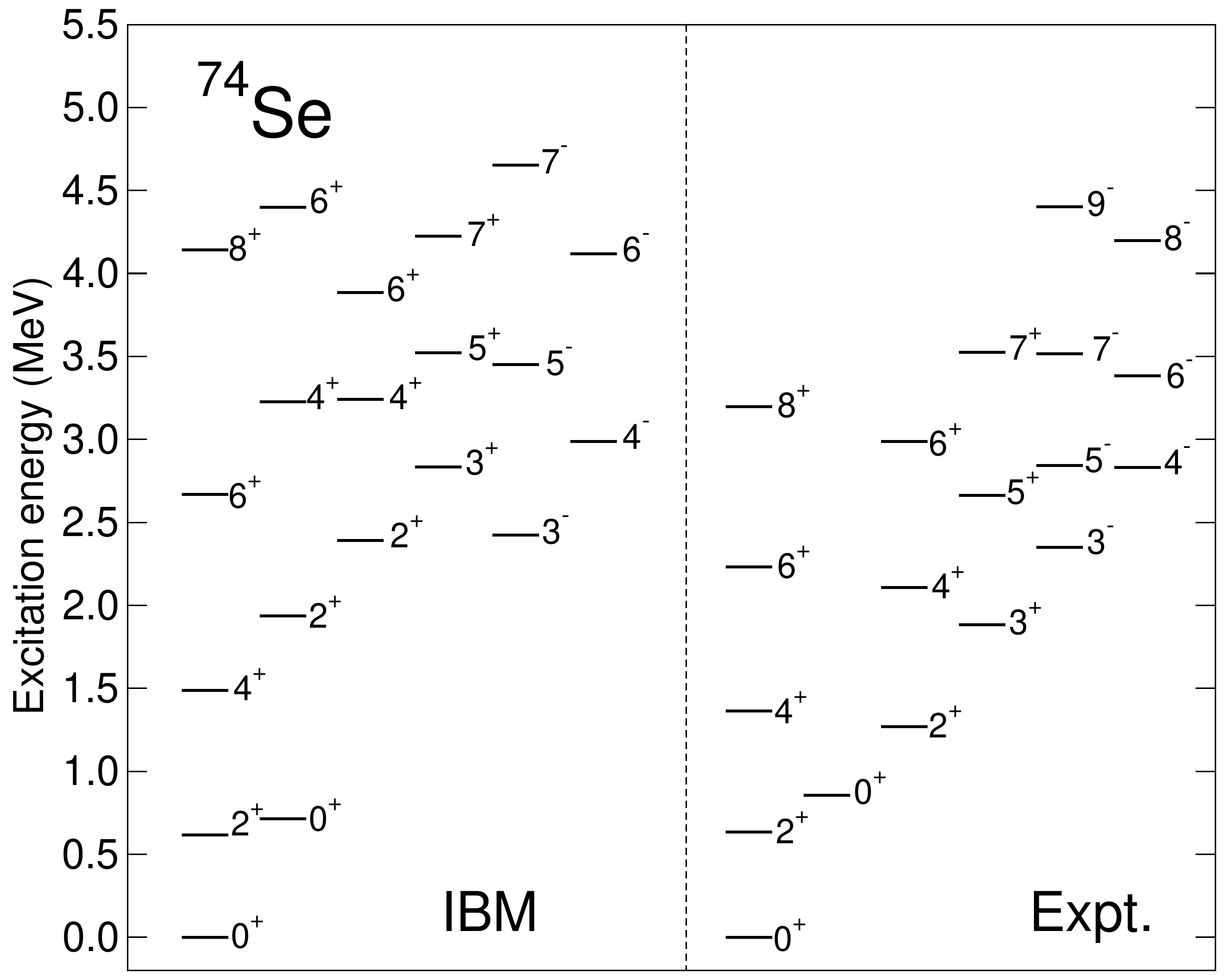}
\caption{Same as Fig.~\ref{fig:ge72}, but for $^{74}$Se.}
\label{fig:se74}
\end{center}
\end{figure}

\begin{figure}
\begin{center}
\includegraphics[width=\linewidth]{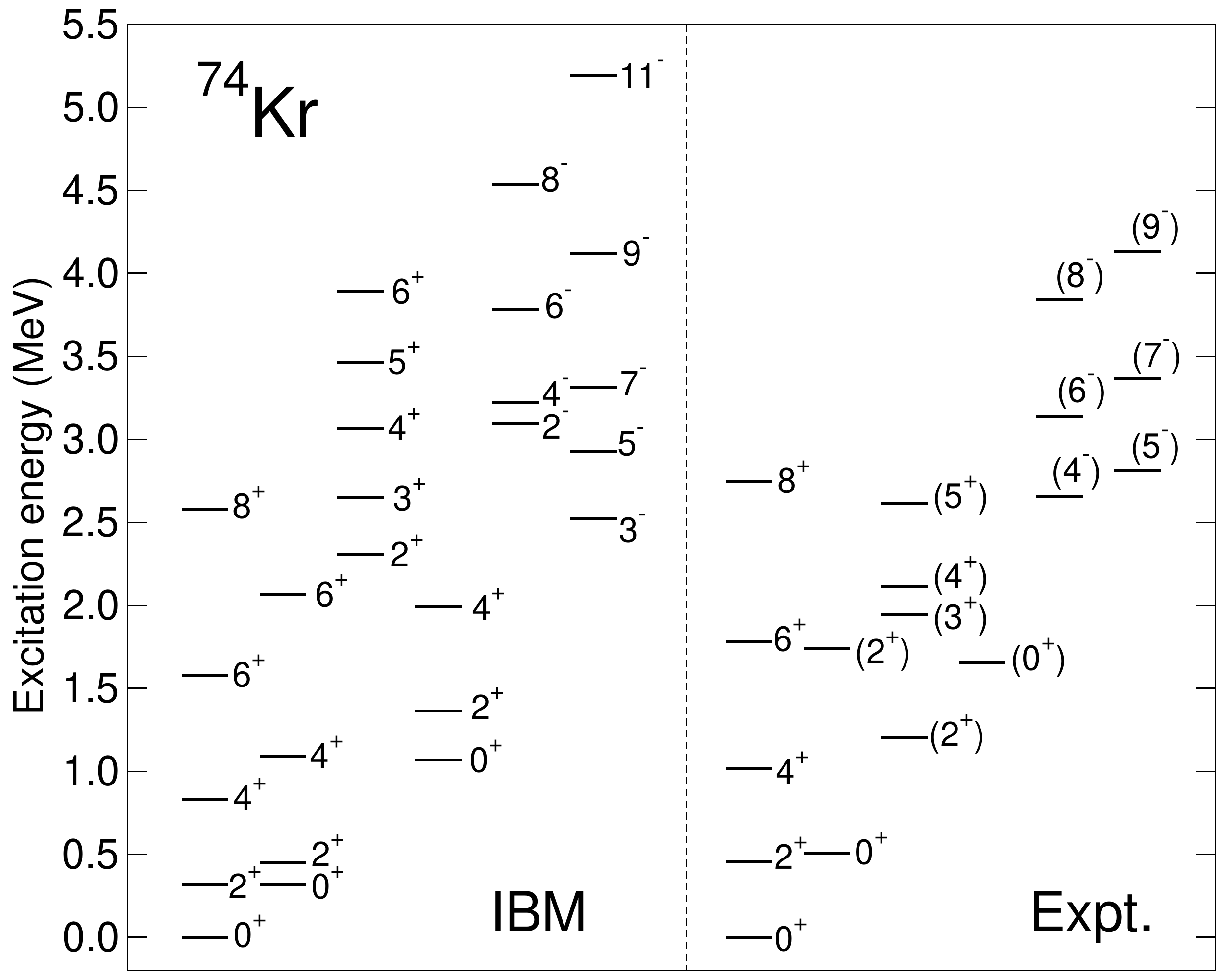}
\caption{Same as Fig.~\ref{fig:ge72}, but for $^{74}$Kr.}
\label{fig:kr74}
\end{center}
\end{figure}

\begin{figure}
\begin{center}
\includegraphics[width=\linewidth]{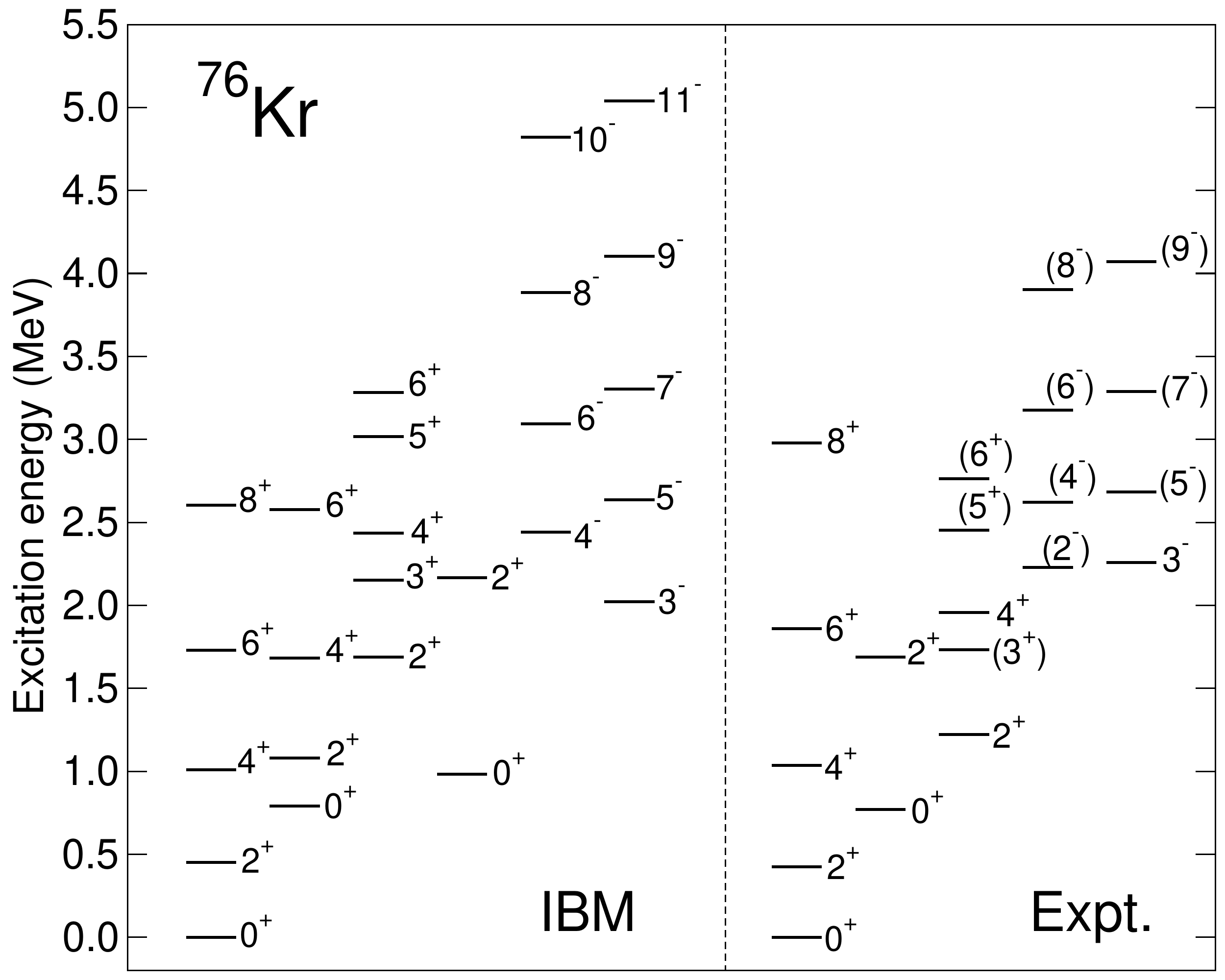}
\caption{Same as Fig.~\ref{fig:ge72}, but for $^{76}$Kr.}
\label{fig:kr76}
\end{center}
\end{figure}

For each nucleus, low-energy band structure 
is studied by organizing states according 
to the structures of the wave functions, some of which 
are shown in Table~\ref{tab:frac}, and to the 
dominant $E2$ transitions within bands. 
The detailed level schemes thus established 
are shown in Figs.~\ref{fig:ge72}-\ref{fig:kr76}, 
for the four studied nuclei. 

As shown in Table~\ref{tab:frac}, 
the spherical normal configuration $[n_0]$ and the 
oblate 2p-2h configuration $[n_1]$ are substantially mixed in the 
ground state $0^+_1$ of $^{72}$Ge. 
The ground-state band with spin $I^\pi \geqslant 2^+$ 
is, however, mostly of $[n_1]$ oblate nature. 
The calculation gives the $0^+_2$ band with 
the bandhead close in energy to the $2^+_1$ level 
(see Fig.~\ref{fig:ge72}). 
Furthermore, a quasi-$\gamma$ band, consisting of the 
$2^+_2$, $3^+_1$, $4^{+}_2$, $5^+_1$ states is obtained. 
The calculated energy level of the $0^+_3$ 
state can be compared to the measured one at the excitation 
energy $E_x=2065$ keV. 
The $0^+_3$ state is here suggested to be purely 
of the 2p-2h oblate nature (99.99 \%). 
The $I^\pi=3^-_1$ state, calculated 
at $E_x=2450$ keV, is the lowest-energy 
negative-parity state, and corresponds to the 
measured one at $E_x=2515$ keV \cite{data}. 
The predicted odd-spin $\Delta I=2$ negative-parity band 
is, however, rather stretched, in comparison with the data, 
due to the level repulsion. 
In addition, the $2^-_1$ state is calculated at 
$E_x=3924$ keV, while the 
experimental one is at $E_x=3036$ keV. 

For $^{74}$Se, the calculation gives the low-lying 
$0^+_2$ state slightly above the $2^+_1$ one 
(Fig.~\ref{fig:se74}). 
The observed even- and odd-spin $\Delta I=2$ positive-parity 
non-yrast bands with bandhead $2^+$ and $3^+$ states 
are here considerably overestimated. 
In this calculation, these non-yrast 
bands are supposed to form a quasi-$\gamma$ band, and 
are mainly accounted for by the $[n_1]$ oblate 
configuration. 
The discrepancy with the data implies that the 
triaxiality is not sufficiently taken 
into account in the IBM system. 
Indeed, as shown in Fig.~\ref{fig:pes-bg}, 
the mapped triaxial quadrupole energy surface for $^{74}$Se 
has an oblate ground-state minimum that is rather rigid 
in the $\gamma$ deformation, while the corresponding 
SCMF energy surface is softer. 
As a consequence, the mapped IBM gives 
a rotational-like energy spectrum like that shown in 
Fig.~\ref{fig:se74}, in which 
the $\gamma$ bandhead is calculated to be  
high with respect to the ground-state band. 

The low-spin positive-parity states 
of $^{74}$Kr are here determined by the strong 
admixture of the three configurations 
(cf. Table~\ref{tab:frac}), 
with the largest contribution 
from the 2p-2h oblate one. 
In the ground-state band, 
the 4p-4h prolate configuration comes 
to play a dominant role with 
increasing spin $I^\pi\geqslant 2^+$, and 
already at $I^\pi=4^+_1$ the $[n_2]$ configuration 
constitutes 94 \% of the wave function. 
As shown in Fig.~\ref{fig:kr74}, the calculation gives 
the low-energy $0^+_2$ band, with the bandhead $0^+_2$ 
level below the $2^+_1$ one. 
The $0^+_2$ state is here predicted to be mainly made 
of the strongly deformed prolate configuration. 
The $2^+_2$ state is interpreted as the first excited 
state of the predicted $0^+_2$ band, and lies close in 
energy to the bandhead $0^+_2$. 
The data, however, suggest a larger energy gap between 
the $0^+_2$ and $2^+_3$ levels. 
The proposed quasi-$\gamma$ band, which is built 
on top of the $2^+_4$ state, is found at much higher 
excitation energy than the observed one. 
One finds a nearly harmonic pattern in the calculated 
quasi-$\gamma$ band. 
This contradicts the data, which 
suggest a pattern typical of a $\gamma$-unstable 
rotor \cite{gsoft}, with the $3^+$ level lying close 
to the $4^+$ one. 
The present calculation further produces another 
rotational-like band built upon the $0^+_3$ state 
at $E_x=1069$ keV. 
The corresponding $0^+_3$ level has 
been observed experimentally at $E_x=1654$ keV. 
The $3^-_1$ state is here calculated to be the lowest 
negative-parity state, but it is not observed experimentally. 
The observed even- and odd-spin negative-parity states 
are the lowest-spin members of the rotational bands \cite{dobon2005}
that extend up to $I=32^-$ and $I=35^-$, respectively.

Figure~\ref{fig:kr76} compares 
between the calculated and experimental 
low-energy excitation spectra for $^{76}$Kr. 
Notable features found in the calculated energy level scheme 
are the low-energy $0^+_2$ and $0^+_3$ excited states, 
both appearing below the $4^+_1$ level. 
The $0^+_2$ band obtained from the calculation 
looks like a rotational band, while in the $0^+_3$  
one the energy gap between the $0^+_3$ bandhead and the 
first excited state in the band, $2^+_4$, is larger than 
in the $0^+_2$ band. 
Experimentally, a $0^+_2$ band with the bandhead 
$0^+_2$ level at $E_x=770$ keV is observed. 
The present calculation suggests an approximate odd-even-spin 
staggering in the quasi-$\gamma$ band; 
that is, the energy levels of 
the odd- and even-spin members of the band higher than $I^\pi=2^+$ 
are close to each other. This is consistent, both qualitatively 
and quantitatively, with the observed quasi-$\gamma$ band. 
Two calculated negative-parity $\Delta I=2$ bands, 
starting from the $3^-_1$ and $4^-_1$ levels, can 
be compared with the observed $3^-$ and $2^-$ bands, 
respectively, which extend up to the spin $I^\pi=19^-$ 
and $22^-$ states \cite{gross1989}. 
At variance with the data, 
the $2^-_1$ level is calculated to be $E_x=2939$ keV, 
which is above the $4^-_1$ one. 
The discrepancy indicates that 
there may be a missing component within the model, e.g., 
the dipole, $p$, boson degrees of freedom. 
However, the inclusion of the $p$ bosons in the 
considered boson model space is practically 
so complicated that it is not pursued in this work.

\subsection{Transition properties}

\begin{table*}
\caption{
\label{tab:em}
Calculated and experimental $B(E2)$, $B(E3)$, and $B(E1)$
transition probabilities (in W.u.), $\rho^2(E0)$ values, 
and spectroscopic quadrupole moments $Q^{(s)}$ (in $e$b). 
The experimental data are adopted from 
Refs.~\cite{data,kibedi2005,stone2005,clement2007}. 
}
 \begin{center}
 \begin{ruledtabular}
  \begin{tabular}{lcccccccc}
 & \multicolumn{2}{c}{$^{72}$Ge} & \multicolumn{2}{c}{$^{74}$Se} 
 & \multicolumn{2}{c}{$^{74}$Kr} & \multicolumn{2}{c}{$^{76}$Kr}\\
\cline{2-3}\cline{4-5}\cline{6-7}\cline{8-9}
 & Calc. & Expt. & Calc. & Expt. & Calc. & Expt. & Calc. & Expt. \\
\hline
$B(E2;2^+_1\rightarrow 0^+_1)$  & 23.5 & $23.5\pm0.4$ & 42.0 & $42.0\pm0.6$ & 66 & $66\pm1$ & 75 & $75\pm1$ \\
$B(E2;4^+_1\rightarrow 2^+_1)$  & 42 & $37\pm5$ & 79 & $80\pm4$ & 107 & $154\pm5$ & 135 & $129\pm3$\\
$B(E2;6^+_1\rightarrow 4^+_1)$  & 48 & $37^{+21}_{-37}$ & 90 & $72\pm 15$ & 147 & $163\pm 16$ & 155 & $146^{+16}_{-5}$\\
$B(E2;0^+_2\rightarrow 2^+_1)$  & 29 & $89\pm2$ & 63 & $77\pm7$ & 83 & $255\pm25$ & 73 & $126^{+6}_{-5}$ \\
$B(E2;2^+_2\rightarrow 2^+_1)$  & 33 & $62^{+9}_{-11}$ & 6.8 & $48\pm14$ & 94 & $25\pm4$ & 35 & $1$ \\
$B(E2;2^+_2\rightarrow 0^+_1)$  & 0.57 & $0.130^{+0.018}_{-0.024}$ & 4.1 & $0.80\pm 0.23$  & 2.0 & $4.3\pm0.5$ & 1.9 & 4 \\
$B(E2;2^+_2\rightarrow 0^+_2)$  & 0.060 & $0.030^{+0.005}_{-0.006}$ & 22 &  & 65 & $24\pm4$ & 32 & 157 \\
[1.0ex]
$B(E3;3^-_1\rightarrow 0^+_1)$  & 29 & $29\pm6$ & 23 & $9\pm2$ & 30 & & 33 & \\
$B(E3;3^-_1\rightarrow 0^+_2)$  & 0.27 & & 0.44 & & 25 & & 0.0013 & \\
$B(E3;3^-_1\rightarrow 2^+_1)$  & 15 & & 14 & & 49 & & 12 & \\
$B(E3;3^-_1\rightarrow 2^+_2)$  & 0.0084 & & 0.45 & & 13 & & 4.5 & \\
$B(E3;5^-_1\rightarrow 2^+_1)$  & 21 & & 15 & & 29 & & 15 & \\
[1.0ex]
$B(E1;3^-_1\rightarrow 2^+_1) \times 10^5$ & 12 & & 0.115 & $0.115 \pm 0.021$ & 36 & & 8.5 & \\
$B(E1;3^-_1\rightarrow 2^+_2)\times 10^{5}$ & 26 & & 13 & & 5.8 & & 5.7 & \\
$B(E1;3^-_1\rightarrow 4^+_1)\times 10^{5}$ & 2.0 & $4.8 \pm 1.0$ & 5.6 & $0.38 \pm 0.10$ & 9.3 & & 1.5 & \\
$B(E1;5^-_1\rightarrow 4^+_1)\times 10^{5}$ & 134 & & 96 & & 341 & & 104 & \\
[1.0ex]
$\rho^2(E0;0^+_2\rightarrow 0^+_1) \times 10^3$ & 9.17 & $9.18 \pm 0.02$ & 23.2 & $23.1 \pm 2.2$ & 113 & $113\pm27$ & 51 & ${}$ \\
$\rho^2(E0;0^+_3\rightarrow 0^+_1) \times 10^3$ & 0.00031 & & 0.35 & & 28 & & 0.19 & \\
$\rho^2(E0;0^+_3\rightarrow 0^+_2) \times 10^3$ & 0.00026 & & 0.41 & & 43 & & 122 & \\
$\rho^2(E0;2^+_2\rightarrow 2^+_1) \times 10^3$ & 0.047 & & 4.9 & & 107 & & 6.0 & \\
[1.0ex]
$Q^{(s)}(2^+_1)$ & 0.22 & $-0.13\pm0.06$ & 0.59 & $-0.36\pm0.07$ & $-0.33$ & $-0.53^{+0.24}_{-0.23}$ & 0.70 & $-0.7\pm0.2$ \\
$Q^{(s)}(2^+_2)$ & $-0.18$ & & 0.046 & & 0.094 & $0.24^{+0.21}_{-0.17}$ & $-0.92$ & $-0.7\pm0.3$ \\
$Q^{(s)}(2^+_3)$ & $0.044$ & & $-0.46$ & & 0.60 & $0.3^{+0.9}_{-0.3}$ & $-0.45$ & $1.0\pm0.4$ \\
  \end{tabular}
 \end{ruledtabular}
 \end{center}
\end{table*}

Electric monopole ($E0$), dipole $(E1)$, quadrupole ($E2$), 
and octupole ($E3$) transition properties are calculated 
with the $E\lambda$ ($\lambda=$ 0, 1, 2, and 3) 
transition operator that is given by
\begin{align}
 \hat T^{E\lambda}=\sum_k {\mathcal{P}_k} 
\hat T^{E\lambda}_k {\mathcal{P}_k} \; .
\end{align}
The transition operator $\hat T^{E\lambda}_k$ 
for each unperturbed space is defined as
\begin{align*}
 & \hat T^{E0}_k = e_{0,k} ( \hat n_d + \hat n_f ) \; , \quad
 \hat T^{E1}_k = e_{1,k} \hat D \; ,
\\
 & \hat T^{E2}_k = e_{2,k} \hat Q \; , \quad
  \hat T^{E3}_k = e_{3,k} \hat O \; .
\end{align*}
$\hat D = ( d^\+ \times \tilde f + f^\+ \times \tilde d )^{(1)}$ 
denotes dipole operator, 
while $\hat n_d$, $\hat n_f$, $\hat Q$, 
and $\hat O$ were already defined. 
For each $E\lambda$ transition, the effective charges 
$e_{\lambda,k}$ are assumed to be the same for all the 
configurations included, 
$e_{\lambda,0}=e_{\lambda,1}=e_{\lambda,2} \equiv e_\lambda$.

Table~\ref{tab:em} lists 
the calculated $B(E2)$, $B(E3)$, and $B(E1)$ 
transition probabilities, 
$\rho^2(E0)=(Z/R_0^2)^2 B(E0)$ values, 
and the spectroscopic quadrupole moments $Q^{(s)}$ 
for low-lying states. 
Here the $E0$ and $E2$ boson charges 
are adjusted, for each nucleus, to reproduce 
the experimental $B(E2;2^+_1\to 0^+_1)$ 
and $\rho^2(E0;0^+_2 \to 0^+_1)$ values, respectively. 
For $^{76}$Kr, since the corresponding $E0$ datum 
is not available, the same $e_0$ value as the one 
for $^{74}$Kr is used.  
The adopted $e_2$ ($e_0$) values 
for $^{72}$Ge, $^{74}$Se, $^{74}$Kr, and $^{76}$Kr 
are, respectively, 0.052 (0.078), 0.061 (0.12), 0.056 (0.18), 
and 0.077 (0.18) $e$b (fm). 
The $E1$ and $E3$ boson charges, 
$e_1=0.036$ $e$b$^{1/2}$ and $e_3=0.042$ $e$b$^{3/2}$, are 
fitted to reproduce the experimental 
$B(E1;3^-_1\to 2^+_1)$ and 
$B(E3;3^-_1\to 0^+_1)$ values for $^{72}$Ge 
and $^{74}$Se, respectively, and are kept constant 
for all four nuclei.

The present calculation yields 
the interband transition rates, including 
$B(E2;0^+_2 \to 2^+_1)$ and $B(E2;2^+_2 \to 2^+_1)$ ones, 
larger than or of the same order 
of magnitude as that for the $B(E2; 2^+_1 \to 0^+_1)$ one. 
The large interband $E2$ transitions are a consequence 
of the configuration mixing. 
The $B(E3;3^-_1 \to 0^+_1)$ rate is often considered 
a signature of the octupolarity, 
and is here calculated to be in the range 20-30 W.u., 
with the fixed $e_3$ boson charge. 
The $3^-_1 \to 0^+_2$ $E3$ transition is orders 
of magnitude weaker than the $3^-_1 \to 0^+_1$ one 
for $^{72}$Ge, $^{74}$Se, and $^{76}$Kr, 
while it is of the same order of 
magnitude as the $3^-_1 \to 0^+_1$ transition in $^{74}$Kr. 
This further confirms the strong configuration 
mixing in the $0^+_2$ state for $^{74}$Kr. 
The model further provides the $E1$ 
properties, and particularly large $B(E1;3^-_1\to 2^+_1)$ 
and $B(E1;5^-_1\to 4^+_1)$ values are obtained 
for $^{74}$Kr. 
The $E0$ transition serves as an empirical signature of 
shape coexistence, and the nuclei in this mass region are 
expected to show pronounced $E0$ transitions, especially, 
between the low-lying $0^+$ states. 
For $^{74}$Kr and $^{76}$Kr, the $E0$ transitions between the 
$0^+_3$ and $0^+_2$ states are also as strong as that between 
the $0^+_1$ and $0^+_2$ ones.

The positive $Q^{(s)}(2^+_1)$ values 
obtained for $^{72}$Ge and $^{74}$Se imply an oblate deformation, 
but are at variance with the experimental values \cite{clement2007}, 
which are negative 
and hence suggest prolate nature for both nuclei. 
The wrong sign of the calculated $Q^{(s)}(2^+_1)$ 
is due to the fact that 
the $2^+_1$ states of both $^{72}$Ge and $^{74}$Se are 
predominantly determined by the 2p-2h oblate configuration 
(see Table~\ref{tab:frac}), since the SCMF calculations 
with the chosen EDF give a pronounced oblate minimum. 
The positive $Q^{(s)}(2^+_2)$ value for $^{72}$Ge 
also indicates the oblate nature of the $2^+_2$ state. 
For $^{74}$Se, on the other hand, the $2^+_2$ state is dominated 
by the 0p-0h nearly spherical configuration, and the 
quadrupole moment is calculated to be rather small 
in magnitude, $Q^{(s)}(2^+_2)<0.1$. 
Similar conclusions on the $Q^{(s)}$ moments have been 
drawn, both quantitatively and qualitatively, from the 
previous configuration-mixing $sd$-IBM studies on 
the Ge and Se isotopes, using the Gogny-type 
EDF \cite{nomura2017ge}. The Gogny-EDF calculation 
carried out in that reference also predicted 
the oblate global minimum in the triaxial quadrupole 
energy surfaces for both $^{72}$Ge and $^{74}$Se.

The predicted $Q^{(s)}(2^+_1)$ and $Q^{(s)}(2^+_2)$
for $^{74}$Kr are within uncertainties of the observed 
values \cite{clement2007} and the sign is 
consistent with the data. 
The negative $Q^{(s)}(2^+_1)$ value reflects the fact that 
the $2^+_1$ state is composed mainly of the 4p-4h 
prolate configuration (cf. Table~\ref{tab:frac}). 
In the $2^+_2$ wave function of the same nucleus, the 
oblate 0p-0h and 2p-2h configurations are more dominant, 
hence the corresponding quadrupole moment has 
positive sign. 
For $^{76}$Kr, the model yields $Q^{(s)}(2^+_1)>0$, 
which indicates oblate ground state but contradicts the data. 
Note that the $2^+_1$ state for $^{76}$Kr is mainly 
accounted for by the 2p-2h 
oblate configuration, while the prolate 4p-4h configuration 
plays only a minor role. 
The negative quadrupole moment 
$Q^{(s)}(2^+_2)=-0.92$ $e$b, obtained by the present 
calculation, is consistent with the experimental value.

\begin{table*}
\caption{
\label{tab:em-cm}
$B(E2)$, $B(E1)$, $B(E3)$, and 
$X(E0/E2)$ ratios, calculated by the $sdf$-IBM with (``IBM'') 
and without 
(``+CM'') the configuration mixing. The experimental data 
are obtained from Refs.~\cite{data,kibedi2005,stone2005}.
}
 \begin{center}
 \begin{ruledtabular}
  \begin{tabular}{lcccccccccccc}
 & \multicolumn{3}{c}{$^{72}$Ge} & \multicolumn{3}{c}{$^{74}$Se} 
 & \multicolumn{3}{c}{$^{74}$Kr} & \multicolumn{3}{c}{$^{76}$Kr}\\
\cline{2-4}\cline{5-7}\cline{8-10}\cline{11-13}
 & IBM & +CM & Expt. & IBM & +CM & Expt. 
 & IBM & +CM & Expt. & IBM & +CM & Expt. \\
\hline
$\frac{B(E2;4^+_1\rightarrow 2^+_1)}{B(E2;2^+_1\rightarrow 0^+_1)}$ 
 & 1.60 & 1.77 & $1.57 \pm 0.21$ & 1.45 & 1.88 & $1.90 \pm 0.10$ 
 & 1.46 & 1.62 & $2.33 \pm 0.08$ & 1.70 & 1.80 & $1.72 \pm 0.05$ \\
[1.0ex]
$\frac{B(E2;0^+_2\rightarrow 2^+_1)}{B(E2;2^+_1\rightarrow 0^+_1)}$ 
 & 1.00 & 1.22 & $3.79 \pm 0.11$ & 0.27 & 1.50 & $1.83 \pm 0.17$ 
 & 0.24 & 1.25 & $3.86 \pm 0.38$ & 1.20 & 0.98 & $1.68^{+0.08}_{-0.07}$ \\
[1.0ex]
$\frac{B(E2;2^+_2\rightarrow 0^+_2)}{B(E2;2^+_1\rightarrow 0^+_1)}$ 
 & 0.14 & 0.003 & $0.0013 \pm 0.0002$ & 0.30 & 0.52 & ${}$ 
 & 0.34 & 0.98 & $0.36 \pm 0.06$ & 0.28 & 0.43 & $2.09 \pm 0.03$ \\
[1.0ex]
$\frac{B(E2;2^+_2\rightarrow 0^+_2)}{B(E2;2^+_2\rightarrow 0^+_1)}$ 
 & 2.7 & 0.11 & $0.23^{+0.05}_{-0.06}$ & 16 & 5.4 & ${}$ 
 & 36 & 33 & $5.58 \pm 1.13$ & 8.3 & 17 & $39$ \\
[1.0ex]
$\frac{B(E1;3^-_1\rightarrow 4^+_1)}{B(E1;3^-_1\rightarrow 2^+_1)}$ 
 & 0.018 & 0.17 & ${}$ & 0.51 & 49 & ${}$ 
 & 3.3 & 0.26 & ${}$ & 0.064 & 0.18 & ${}$ \\
[1.0ex]
$\frac{B(E3;3^-_1\rightarrow 2^+_1)}{B(E3;3^-_1\rightarrow 0^+_1)}$ 
 & 0.057 & 0.50 & ${}$ & 0.015 & 0.59 & ${}$ 
 & 0.61 & 1.65 & ${}$ & 0.036 & 0.35 & ${}$ \\
$\frac{10^3 \rho^2(E0;0^+_2\rightarrow 0^+_1) e^2 R_0^4}{B(E2;0^+_2\rightarrow 2^+_1)}$ 
 & 37 & 11 & ${}$ & 444 & 13 & $17 \pm 4$ 
 & 130 & 48 & $10.5 \pm 4.8$ & 211 & 24 & ${}$ \\
  \end{tabular}
 \end{ruledtabular}
 \end{center}
\end{table*}

The configuration mixing has an influence on 
qualitative features of 
some transition properties of the low-lying states, 
especially those that involve 
the excited $0^+$ states. 
Table~\ref{tab:em-cm} lists the ratios 
of $B(E2)$, $B(E1)$, and $B(E3)$ transition probabilities, 
and the $E0/E2$ ratio defined as 
$X(E0/E2) \equiv {\rho^2(E0;0^+_2 \to 0^+_1)e^2 R_0^4}/{B(E2;0^+_2 \to 2^+_1)}$, that are obtained from the $sdf$-IBM calculations 
with and without the configuration mixing. 
The configuration mixing does not have much 
influence on the 
$\frac{B(E2;4^+_1\rightarrow 2^+_1)}{B(E2;2^+_1\rightarrow 0^+_1)}$
ratio of the inband transitions, but increases the 
$\frac{B(E2;0^+_2\rightarrow 2^+_1)}{B(E2;2^+_1\rightarrow 0^+_1)}$ 
one for $^{74}$Se and $^{74}$Kr by a factor of 5. 
Both the 
$\frac{B(E1;3^-_1\rightarrow 4^+_1)}{B(E1;3^-_1\rightarrow 2^+_1)}$ 
and 
$\frac{B(E3;3^-_1\rightarrow 2^+_1)}{B(E3;3^-_1\rightarrow 0^+_1)}$ 
ratios are generally increased by orders of magnitude. 
The $X(E0/E2)$ ratio is significantly reduced 
for most of the nuclei, and becomes more 
consistent with the data \cite{kibedi2005} after the 
mixing.

\begin{figure}
\begin{center}
\includegraphics[width=\linewidth]{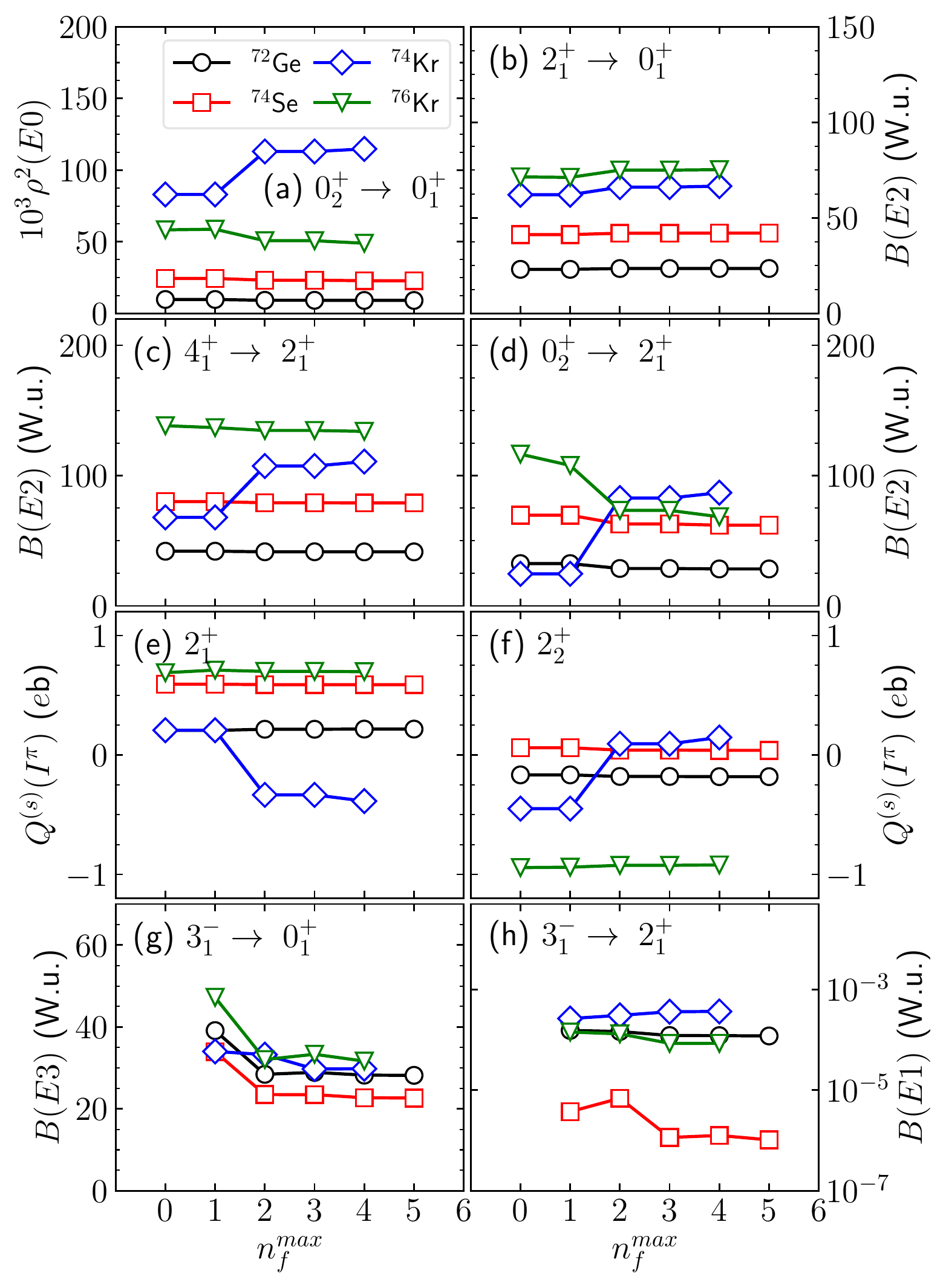}
\caption{Calculated transition properties of 
$^{72}$Ge, $^{74}$Se, $^{74}$Kr, and $^{76}$Kr as functions of 
the maximum number of $f$ bosons, $n^\text{max}_f$: 
(a) $\rho^2(E0;0^+_2\to 0^+_1)$, (b) $B(E2;2^+_1 \to 0^+_1)$, 
(c) $B(E2;4^+_1 \to 2^+_1)$, (d) $B(E2;0^+_2 \to 2^+_1)$, 
(e) $Q^{(s)}(2^+_1)$, (f) $Q^{(s)}(2^+_2)$, 
(g) $B(E3;3^-_1 \to 0^+_1)$, and (h) $B(E1;3^-_1 \to 2^+_1)$. 
The same boson effective charges are used for each set of 
the calculations corresponding to different $n^\text{max}_f$.}
\label{fig:nf-tr}
\end{center}
\end{figure}

\subsection{Sensitivity to the truncation of the $f$ boson number\label{sec:fboson}}

The results presented so far have been 
obtained with the maximum 
number of $f$ bosons, $n^\text{max}_f$, being 
truncated as $n^\text{max}_f=3$ for each $[n_k]$ subspace. 
To confirm that this truncation is adequate, 
calculated $\rho^2(E0)$ values, $B(E2)$, $B(E3)$, 
and $B(E1)$ transition rates, and spectroscopic 
quadrupole moments $Q^{(s)}$ are shown 
in Fig.~\ref{fig:nf-tr} as functions of 
$n^\text{max}_f$. 
One realizes that the 
calculated quantities generally become stable 
for $n^\text{max}_f \geqslant 2$. This, in turn, indicates 
that more than one $f$ boson should be needed in the 
considered boson model space to obtain converged results. 

It is, nevertheless, worth noting that many of 
the transition properties calculated for $^{74}$Kr change 
significantly from $n^\text{max}_f=1$ to 2, 
an illustrative example being 
the $Q^{(s)}(2^+_1)$ and $Q^{(s)}(2^+_2)$ moments 
for which the sign changes 
at $n^\text{max}_f=2$. 
Since for $^{74}$Kr the 4p-4h prolate components 
make large contributions to the low-lying states, 
the structure of the $2^+_1$ 
wave function is supposed to be sensitive 
to the increasing $n^\text{max}_f$: 
For $n^\text{max}_f=1$ ($n^\text{max}_f=2$), 
the three configurations $[n_0]$, $[n_1]$, and $[n_2]$ 
constitute 12 (28) \%, 24 (45) \%, and 63 (27) \% 
of the $2^+_1$ wave function of $^{74}$Kr. 
A similar argument holds for the $2^+_2$ state. 

On the other hand, the truncation for 
$n^\text{max}_f$ does not have 
as noticeable impacts on the excitation energies as in 
the case of the transition properties, and hence 
these are not discussed here.

\section{Concluding remarks\label{sec:summary}}

Based on the 
nuclear density functional theory, 
configuration mixing and octupole degrees of freedom 
are incorporated microscopically in the 
interacting boson model. 
The constrained SCMF calculations 
using the universal EDF and pairing interaction 
provide two sets of the potential energy surfaces, 
one with the triaxial quadrupole, and the other 
with the axially symmetric 
quadrupole and octupole degrees of freedom. 
A model Hamiltonian for the configuration mixing 
$sdf$-IBM is determined by mapping the SCMF energy 
surfaces onto the bosonic counterparts.

In the illustrative application to the transitional 
nuclei $^{72}$Ge, $^{74}$Se, $^{74}$Kr, and $^{76}$Kr, 
the configuration mixing of the normal and intruder states 
in the  $sdf$-IBM significantly lowers 
the $0^+_2$ energy level. 
The predicted low-energy positive-parity 
states are characterized by the strong admixture of 
the nearly spherical, weakly deformed oblate, and strongly 
deformed prolate shapes. 
In particular, the prolate deformation 
plays an important role in the ground state of $^{74}$Kr. 
The configuration mixing has influences on 
the transition properties, as well as the excitation 
energies. 
Notable examples are the increase 
of the $B(E2;0^+_2 \to 2^+_1)$ transition rate 
and the reduction of the $X(E0/E2)$ ratio after the mixing, 
both of which quantities are signatures of shape 
coexistence and mixing. 
For the $N=40$ isotones, on the other hand, 
the model calculation predicts the spectroscopic 
quadrupole moments of the $2^+_1$ state with 
positive sign, which contradicts experiment. 
The discrepancy arises from the fact that large 
amounts of the oblate components are 
contained in the corresponding $2^+_1$ wave functions. 
The nature of the IBM wave functions reflects 
the topology of the SCMF potential energy surfaces, 
which indeed suggest the pronounced oblate 
minimum in these nuclei. 
The mapped Hamiltonian produces 
negative-parity bands with the lowest-energy state 
with $I^\pi=3^-_1$ at the excitation energy 
$E_x\approx 2$ MeV, and the electric octupole 
transitions to the ground state. 
The low-energy negative-parity 
states are here shown to be made predominantly of 
the deformed intruder configurations.

The proposed model can be used further for extensive 
studies on neutron-rich nuclei, in which both shape 
coexistence and octupole collectivity are expected 
to play an important role at low energy, and which 
are experimentally of much interest. 

\acknowledgments
This work is financed within the Tenure Track Pilot Programme of 
the Croatian Science Foundation and the \'Ecole Polytechnique 
F\'ed\'erale de Lausanne, and the Project TTP-2018-07-3554 Exotic 
Nuclear Structure and Dynamics, with funds of the Croatian-Swiss 
Research Programme.

\appendix

\section{Formulas for the bosonic potential energy surface\label{sec:pes}}

In the diagonal element of the matrix 
${\mathbf{E}}(\beta_2,\gamma,\beta_3)$, 
Eq.~(\ref{eq:pes-diag}), the expectation value of the 
unperturbed Hamiltonian reads
\begin{widetext}
\begin{align}
\label{eq:pes-diag2}
&
\braket{\Psi_k(\vec\alpha_k)|\hat H_k|\Psi_k(\vec\alpha_k)}
=
\frac{n_k (\tilde\epsilon_{s,k} + \tilde\epsilon_{d,k} \beta_{2,k}^{2} + \tilde\epsilon_{f,k} \beta_{3,k}^{2})}
{1+\beta_{2,k}^{2}+\beta_{3,k}^{2}}
+\frac{n_k(n_k-1)}{(1+\beta_{2,k}^2+\beta_{3,k}^2)^2}
\nonumber\\
&
\quad\quad\quad\quad
\times
\Biggl[
\kappa_{2,k}
\biggl[
4\beta_{2,k}^2
- 4\tilde{\chi}_{k} \beta_{2,k}^3 \cos{3\gamma} 
+ \tilde{\chi}_{k}^2 \beta_{2,k}^4
+ \tilde{\chi}_{k}' \beta_{2,k} \beta_{3,k}^2 
\left\{
4 - \tilde{\chi}_{k} \beta_{2,k} (3\cos{2\gamma}-1)
\right\}
+ \tilde{\chi}_{k}'^2 \beta_{3,k}^4
\biggr]
\nonumber\\
&
\quad\quad\quad\quad\quad\quad
-4\kappa_{3,k}\beta_{3,k}^2
\left\{
1 + 2{\tilde{\chi}}_{k}'' \beta_{2,k}
+ \frac{1}{8}
\beta_{2,k}^2
{\tilde{\chi}}_{k}''^2
\left(
9-\cos2\gamma
\right)
\right\}
\Biggr]
+
\frac{1}{30}\eta_k
\frac{n_k(n_k-1)(n_k-2)}
{(1+\beta_{2,k}^2+\beta_{3,k}^2)^3}
\beta^6_{2,k}\sin^2{3\gamma}
\; .
\end{align}
\end{widetext}
Note that, in the first line of the above expression, 
$\tilde\epsilon_{s,k} = 5\kappa_{2,k} - 7\kappa_{3,k}$, 
$\tilde\epsilon_{d,k} = \epsilon_{d,k} + (1+\chi_k^2)\kappa_{2,k} - 6\rho_k -\frac{7}{5}\chi_k''^2\kappa_{3,k}$, and 
$\tilde\epsilon_{f,k} = \epsilon_{f,k} - \frac{5}{7} {\chi_k'}^2\kappa_{2,k} + (1+{\chi_k''}^2)\kappa_{3,k}$, 
and that 
$\tilde{\chi}_k \equiv \sqrt{\frac{2}{7}} \chi_k$, 
$\tilde{\chi}_k' \equiv \frac{2}{\sqrt{21}} \chi_k'$, and 
$\tilde{\chi}_k'' \equiv -\frac{2}{\sqrt{15}} \chi_k''$. 
Nondiagonal elements of ${\bf E}(\beta_2,\gamma,\beta_3)$, 
Eq.~(\ref{eq:pes-nondiag}), coming from the mixing interaction 
$\hat V_{k,k+1}$, with $k=0$ or 1, are calculated as
\begin{widetext} 
\begin{align}
\label{eq:pes-nondiag2}
 E_{k,k+1}(\beta_2,\gamma,\beta_3)
=&\,E_{k+1,k}(\beta_2,\gamma,\beta_3)
\nonumber\\
=&
\sqrt{(n_k+1)n_{k+1}}
\,\frac{\omega_{s,k} + \omega_{d,k} \beta_{2,k+1}^2 + \omega_{f,k} \beta_{3,k+1}^2}
{1+\beta_{2,k+1}^2+\beta_{3,k+1}^2}
\left[
\frac{1+\beta_{2,k}\beta_{2,k+1}+\beta_{3,k}\beta_{3,k+1}}
{\sqrt{(1+\beta_{2,k}^2+\beta_{3,k}^2)(1+\beta_{2,k+1}^2+\beta_{3,k+1}^2)}}
\right]^{n_k} \; .
\end{align}
\end{widetext}

\bibliography{refs}

\end{document}